\documentclass[12pt]{article}
\usepackage{graphicx}
\usepackage{epsf,epsfig,amsmath,amssymb,verbatim,mathrsfs}
\usepackage{epstopdf}
\def\beq{\begin{equation}}
\def\eeq{\end{equation}}
\def\bea{\begin{eqnarray}}
\def\eea{\end{eqnarray}}
\def\xfb{\, {\rm fb}}
\def\xfbinv{\, {\rm fb}^{-1}}
\def\tev{\, {\rm TeV}}
\def\gev{\, {\rm GeV}}
\newcommand{\gsim}{\lower.7ex\hbox{$\;\stackrel{\textstyle>}{\sim}\;$}}
\newcommand{\lsim}{\lower.7ex\hbox{$\;\stackrel{\textstyle<}{\sim}\;$}}
\begin{document}
\pagestyle{empty}

\vspace{1cm}

\begin{center}
{\Large\bf Emergent Electroweak Symmetry Breaking \\ with Composite $W, Z$ Bosons}

\vspace{1cm}

{\sc Yanou Cui,$^{a,}$}\footnote{E-mail:  ycui@physics.harvard.edu}
{\sc Tony Gherghetta$^{b,}$}\footnote{E-mail:  tgher@unimelb.edu.au}
{\sc\small and} {\sc James D. Wells$^{c,d,}$}\footnote{E-mail:
james.wells@cern.ch}

\vspace{0.5cm}

{\it\small $^a$Jefferson Physical Laboratory, Harvard University,
Cambridge, Massachusetts 02138, USA\\
$^b$School of Physics, University of Melbourne, Victoria 3010, Australia\\
$^c$CERN Theory Group (PH-TH), CH-1211 Geneva 23, Switzerland\\
$^d$MCTP, University of Michigan, Ann Arbor, Michigan 48109, USA}
\end{center}


\vspace{1cm}
\begin{abstract}
We present a model of electroweak symmetry breaking in a warped
extra dimension where electroweak symmetry is broken at the UV
(or Planck) scale. An underlying conformal symmetry is broken at the
IR (or TeV) scale generating masses for the electroweak gauge bosons
without invoking a Higgs mechanism. By the AdS/CFT correspondence
the $W,Z$ bosons are identified as composite states of a strongly-coupled
gauge theory, suggesting that electroweak symmetry breaking is an
emergent phenomenon at the IR scale. The model satisfies electroweak
precision tests with reasonable fits to the $S$ and $T$ parameter. In
particular the $T$ parameter is sufficiently suppressed since the model
naturally admits a custodial $SU(2)$ symmetry. The composite nature of
the $W,Z$-bosons provide a novel possibility of unitarizing $WW$ scattering
via form factor suppression. Constraints from LEP and the Tevatron as well as
discovery opportunities at the LHC are discussed for these composite
electroweak gauge bosons.
\end{abstract}

\newpage
\setcounter{page}{1}
\setcounter{footnote}{0}
\pagestyle{plain}

\tableofcontents
\vfill\eject

\section{Introduction}

A major goal of the Large Hadron Collider (LHC) is to unveil the
origin of electroweak symmetry breaking (EWSB). In the Standard
Model (SM) the Higgs mechanism provides a particularly attractive
way to break electroweak symmetry and generate mass. It is most
simply implemented by introducing the Higgs boson, a scalar field
that spontaneously breaks  the electroweak symmetry by obtaining a
vacuum expectation value (VEV). If the Higgs boson is an elementary
scalar field then its mass can be stabilized by low-energy
supersymmetry. In this case supersymmetry breaking triggers the
breaking of electroweak symmetry. Alternatively the Higgs boson may
be composite and therefore can be stabilized with strong dynamics at
the TeV scale~\cite{Kaplan:1983fs}. A composite Higgs boson model
was recently constructed using the holographic dual of a warped
dimension~\cite{Agashe:2004rs}. But the presence of a Higgs boson is
not obligatory. As is well-known by analogy with QCD, electroweak
symmetry can also be broken by condensates in technicolor. In fact,
via the AdS/CFT correspondence, a technicolor-like Higgsless model
can be constructed in a warped extra dimension providing a more
recent incarnation of this idea~\cite{Csaki:2003zu}. Nevertheless
the underlying feature of all these models is that electroweak
symmetry is broken at the electroweak scale where, in particular,
the Standard Model gauge fields and fermions receive a mass
by coupling to an external Higgs sector. To generate
(not just stabilize) the TeV mass scale or gauge hierarchy in a natural
way via dimensional transmutation, it seems inevitable to have strong
dynamics in this external sector (whether it be associated with electroweak or
supersymmetry breaking), together with the Standard Model gauge fields
and fermions which do not directly partake in the underlying dynamics.

Yet there is another possibility for mass generation. Just
like the hadron mass spectrum in QCD, there is no need to invoke the
Higgs mechanism to generate a mass. Instead the fermions and
$W,Z$-bosons could be composite states which directly obtain a mass
from the underlying confining strong dynamics. This idea is not new~\cite{AGC},
since QCD contains states which mimic the electroweak gauge bosons.
Specifically it was noticed that low-energy QCD can be interpreted
as a spontaneously broken gauge theory, where the SU(2) isospin
triplet $\rho$ meson is the massive gauge field of a hidden local
symmetry~\cite{Bando:1984ej}. Interestingly this interpretation can
effectively explain $\rho$-coupling universality, $\rho$-meson
dominance, and the high-energy $\pi$-$\pi$ scattering cross section.
This bears some resemblance with electroweak symmetry suggesting
that the $W,Z$-bosons might be composite. Furthermore, unlike global
symmetries, gauge symmetries do not lead to new conserved charges
and merely remove the redundancy in our description of massless
spin-1 particles with spacetime vector fields. This has led to the suggestion
that gauge symmetries are not fundamental~\cite{Gross:1997wg}. In
fact this is supported by duality in four-dimensional (4D) supersymmetric theories
which imply that gauge symmetries in the SM or even general
relativity could be long-distance artifacts~\cite{Seiberg:1995ac}.
So the idea that electroweak gauge bosons are composite represents
an intriguing although relatively unexplored possibility.

In this paper we examine this possibility by constructing an EWSB
model with composite $W,Z$ bosons. Since the underlying theory is
inherently strongly-coupled we use the AdS/CFT correspondence
\cite{Maldacena:1997re,ArkaniHamed:2000ds} to construct a 
calculable five-dimensional (5D) model using a warped fifth 
dimension~\cite{Randall:1999ee}. The electroweak symmetry will be 
assumed to be preserved on the IR brane and the
bulk, while it will be broken on the UV brane. Hence in this model
the IR brane is only used to break conformal symmetry and generate
massive states. This is opposite from the original Randall-Sundrum
model and all other warped models where EWSB is assumed to occur on
the IR brane. This implies that in our model electroweak gauge symmetry is
not a fundamental symmetry and merely emerges at the IR scale when
the conformal symmetry is broken. Moreover even though our model is
effectively Higgsless at low energies because EWSB occurs on the UV
brane and the Higgs sector decouples, it  still differs from the
usual technicolor-like Higgsless model~\cite{Csaki:2003zu} where
EWSB occurs at the IR scale and the $W,Z$-boson are elementary
fields.

Brane kinetic terms are a crucial feature of our model.
In order to identify the lightest Kaluza-Klein (KK) states
with composite $W, Z$ bosons they are introduced to separate the
lowest-lying KK mode from the rest of the KK tower. This guarantees
that the heavier KK resonances of the $W,Z$ bosons are sufficiently
heavy to evade direct experimental bounds. This leads to $W$ and $Z$
boson profiles that are effectively localized near the IR brane,
rendering them composite states of the dual 4D theory. While the
main focus of this work will be on the gauge boson sector we will
make simplifying assumptions regarding the fermion sector. Since
electroweak symmetry is broken on the UV brane, fermions must at
least couple to the UV brane and have universal profiles in the bulk
to ensure gauge coupling universality. This will be left for future
work and for simplicity we will assume the fermions are localized on
the IR brane.

The idea of composite weak gauge bosons has been previously explored
in the literature~\cite{AGC} where attempts were made to construct the
underlying preon models based on asymptotically-free QCD.
Unlike these earlier attempts the AdS/CFT dictionary relates our composite
gauge boson model to a dual 4D conformal field theory at large `t Hooft
coupling. In addition the weakly-coupled 5D gravity dual improves
calculability allowing a more quantitative analysis of the composite
model. Consequently, a precision electroweak analysis can be
performed leading to reasonable agreement with the $S, T$ parameters.
In particular, the $T$ parameter is sufficiently suppressed due to a
custodial SU(2) symmetry that naturally occurs in the model. Since
the gauge bosons are composite there are also various interesting
phenomenological aspects to study. In fact due to the strong
coupling the underlying composite nature of the gauge bosons is much
less partonic at moderate Bjorken $x$ compared to QCD hadrons. In
particular the composite nature of the $W,Z$ boson provides a
novel unitarization mechanism for $WW$ scattering based on form
factor suppression, and suggests that deviations in the scattering
amplitude may be measurable, giving rise to a distinctive signal at the
LHC.

The organization of this paper is as follows. In Section 2 we
present an overview of the model, outlining our various assumptions
before presenting the full details of the 5D model in a warped
dimension. The electroweak precision analysis is presented in
Section 3 where it is shown how the model naturally admits a
custodial SU(2) symmetry. The $S$ and $T$ parameter are both shown
to be consistent with electroweak precision tests. Section 4 is
devoted to a preliminary study on $WW$ scattering and unitarity.
Various constraints and signatures of composite weak gauge bosons 
at LEP, Tevatron, and the LHC are discussed in Section 5. Finally in
Appendix A we present an alternative derivation of how brane kinetic
terms give rise to a light KK mode, while in Appendix B we present
a form factor computation using the overlap integral.

\section{Emergent EWSB}
\subsection{The 5D Model}
\label{sec5dmodel}
We begin by defining our EWSB model using the Randall-Sundrum framework~\cite{Randall:1999ee}.
Consider a slice of AdS$_5$ with 5D metric
\beq
ds^2=\frac{1}{(kz)^2}(\eta_{\mu\nu}dx^\mu dx^\nu+dz^2)\equiv g_{MN}dx^Mdx^N~,
\eeq
where $k$ is the AdS curvature scale. The 5D spacetime indices are written
as $M=(\mu,5)$, with $\mu=0,1,2,3$, and $\eta_{\mu\nu}={\rm diag}(-+++)$
is the Minkowski metric. The fifth dimension $z$ is compactified on a $Z_2$
orbifold, with a UV (IR) brane located at the fixed point $z^*=z_{UV}(z_{IR})$.
The $z$ coordinate is related to the 4D energy scale, and
the scale of the UV (IR) brane is chosen to be $z_{UV}=k^{-1},
(z_{IR}={\cal O}({\rm TeV}^{-1}))$ respectively, where
$k\simeq M_P=2.4\times 10^{18}$ GeV is the reduced Planck scale.

The 5D bulk is assumed to have an electroweak symmetry,
$SU(2)_L\times U(1)_Y$, while on the boundaries the electroweak
symmetry is preserved on the IR brane but broken on the UV brane.
This is to ensure that the $W,Z$ bosons are identified with the
lowest-lying KK modes peaking towards the IR brane, so that by the AdS/CFT
correspondence they are interpreted as composite states. On the UV
brane the symmetry is broken by imposing Dirichlet boundary
conditions that realizes the SM symmetry breaking pattern
$SU(2)_L\times U(1)_Y\rightarrow U(1)_Q$. As pointed out in
\cite{Csaki:2003zu} EWSB via Dirichlet boundary conditions can be
more naturally understood as the limit of a Higgs mechanism on the
boundary with a very large VEV. Since the breaking is on the UV
brane, the Higgs sector decouples and the model is effectively
Higgsless at low energies. The unbroken electromagnetic group
$U(1)_Q$ on the UV brane leads to a massless photon. In addition,
even though the strong force is irrelevant for our discussion, there
are massless gluons from the unbroken $SU(3)$ color symmetry. So in
our setup, the massive $W,Z$ gauge fields are mostly composite,
while the massless gauge fields are mostly elementary.

There is an immediate problem with identifying the lowest-lying KK
states with the $W,Z$ bosons. Since in the warped dimension
the KK modes are essentially evenly spaced, the next-heaviest
KK states will have masses at approximately 200 GeV. This obviously
contradicts direct searches and electroweak precision data that require
additional electroweak gauge bosons $(i.e.\, W',Z') $ to be heavier than
about $1\,\rm TeV$~\cite{Amsler:2008zzb}. Hence to obtain a desirable
mass spectrum we will introduce brane-localized kinetic terms~\cite{Georgi:2000ks,
Davoudiasl:2004pw}. We will see that this leads to very light lowest-lying
KK modes for the $W,Z$ bosons, while the remaining heavier KK modes
of the $W,Z$-boson and photon will appear at the TeV scale.
Brane kinetic terms will be added on both branes consistent with the brane
symmetry. This includes mass-dimension $-1$ brane kinetic term coefficients
$\zeta_Q$ for $U(1)_Q$ on the UV brane, and $\zeta_L,\zeta_Y$ for
$SU(2)_L$ and $U(1)_Y$ on the IR brane, respectively. Thus the 5D
action of our model is given by
\bea
S=\int d^4x\,dz\sqrt{-g}\left[-\frac{1}{4}(F^{La}_{MN})^2-\frac{1}{4}(F^{Y}_{MN})^2-\frac{1}{2}(kz)\delta{(z-z_{UV})} \frac{\zeta_Q}{g_{Y5}^2+g^2_{L5}}(g_{Y5}F^{L3}_{\mu\nu}+g_{L5}F^{Y}_{\mu\nu})^2\right. \nonumber \\
\left.-\frac{1}{2}(kz)\delta{(z-z_{IR})}\left(\zeta_L(F^{La}_{\mu\nu})^2+\zeta_Y(F^{Y}_{\mu\nu})^2\right)\right],
\label{5daction}
\eea
where the 5D field strengths $F^L_{MN},F^Y_{MN}$ with associated
gauge fields $A^L_M,B_M$, and 5D gauge couplings $g_{L5},g_{Y5}$,
are for $SU(2)_L, U(1)_Y$, respectively. Since only the UV boundary is
Higgsed, the fifth components of the gauge fields are unphysical~\cite{Csaki:2005vy}.
This ensures that our model contains no $A_5$-like holographic Higgs bosons.

Note that the brane kinetic terms in (\ref{5daction}) are crucial ingredients
in our model. They are always allowed on the branes at tree level by
the breaking of 5D Poincare symmetry. But more
importantly in any 5D theory with bulk gauge fields, as well as
charged bulk matter subject to orbifold boundary
conditions~\cite{Georgi:2000ks} or confined to the
branes~\cite{Dvali:2000rx}, the divergent radiative corrections to
the gauge propagator requires that brane kinetic terms be included as
counterterms. Therefore from the effective field theory perspective
the coefficients of the brane kinetic terms are free parameters of
the theory with their exact values depending on the UV completion.
Although naive dimensional analysis (NDA) estimates the size of the
coefficients to be of order the compactification scale of the fifth
dimension~\cite{Carena:2002me,Carena:2002dz,Chacko:1999hg},
large brane kinetic terms are perturbatively consistent~\cite{Ponton:2001hq}.
Thus, just like previous analyses in \cite{Carena:2002dz, Davoudiasl:2004pw},
we will assume the brane
kinetic term coefficients to be free parameters that are fixed by
experimental data, and do not constrain them to be of their NDA size.

\subsubsection{The gauge boson mass spectrum}

The boundary conditions that realize the symmetry breaking pattern
and include the brane kinetic terms are
\bea
&&z=z_{UV}:  \left\{
\begin{array}{l}
\hbox{$\partial_z(g_{Y5}A^{L3}_\mu+g_{L5}B_\mu)
+\zeta_Q k z_{IR}\Box(g_{Y5}A^{L3}_\mu+g_{L5}B_\mu)=0,$} \\
\hbox{$g_{L5}A^{L3}_\mu-g_{Y5}B_\mu=0,$} \\
\hbox{$A^{L1,2}_\mu=0,$} \label{bcuv}
\end{array}
\right.
\\
&&z=z_{IR}: \left\{
\begin{array}{l}
\hbox{$\partial_z A_\mu^{La}-\zeta_L k z_{IR}\Box A_{\mu}^{La}=0,$} \\
\hbox{$\partial_z B_\mu-\zeta_Y k z_{IR}\Box B_\mu=0,$} \label{bcir}
\end{array}
\right.
\eea
where $\Box=\eta_{\mu\nu}\partial^\mu \partial^\nu$ is the
4D Laplacian. Imposing these boundary conditions on the bulk
solutions will lead to the mass spectrum and is similar to that
performed in Refs.~\cite{Carena:2002dz,Davoudiasl:2002ua,Davoudiasl:2004pw}.
Gauge fields which are mixed by the boundary conditions share the
same KK mass spectrum (although with a different 5D profile for the
same KK mode). In particular this is the KK tower containing both
$A_\mu^{L3}$ and $B_\mu$. When $B_\mu$  is dominant, it is
identified as the KK photon tower, while when  $A_\mu^{L3}$ is dominant, it is
identified as the KK $Z$-boson tower. For the other tower,
$A_\mu^{L\pm}$ is identified as the KK $W$-boson tower where
$A_\mu^{L\pm}$ denotes the linear combination
$\frac{1}{\sqrt{2}}(A_\mu^{L1}\mp iA_\mu^{L2}$).
The KK decomposition is therefore given by
\bea
A_\mu^{L3}(x,z)&=&f_0^{L3}(z)\gamma_\mu(x)+\displaystyle\sum_{n=1}^{\infty}f^{L3}_n(z)Z_\mu^{(n)}(x),
\label{KKdec1}\\
B_\mu(x,z)&=&f^B_0(z)\gamma_\mu(x)+\displaystyle\sum_{n=1}^{\infty}f^B_n(z)Z_\mu^{(n)}(x),
\label{KKdec2}\\
A_\mu^{L\pm}(x,z)&=&\displaystyle\sum_{n=1}^{\infty}f^{L\pm}_n(z)W_\mu^{(n)\pm}(x).
\label{KKdec3}
\eea
We have separated out the photon $\gamma_\mu$ in the decomposition of $A_\mu^{L3}, B_\mu$
since later we will show that there is a massless flat zero mode in this KK tower.
Substituting these decompositions into the boundary conditions (\ref{bcuv}) and (\ref{bcir}), leads to the
explicit boundary conditions for the 5D profile functions:
\bea
&&z=z_{UV}:  \left\{
\begin{array}{l}
\hbox{$\partial_z(g_{Y5}f^{L3}_n(z)+g_{L5}f^{B}_n(z))+\zeta_Q\, k z_{IR}\,m_{Z_n}^2(g_{Y5}f^{L3}_n(z)+g_{L5}f^{B}_n(z))=0,$} \\
   \hbox{$g_{L5}f^{L3}_n(z)-g_{Y5}f^{B}_n(z)=0,$} \\
   \hbox{$f^{L\pm}_n(z)=0,$} \\
     \end{array}\label{bcz0}
\right.
\\
&&z=z_{IR}: \left\{
\begin{array}{l}
   \hbox{$\partial_z(f^{L3}_n(z))-\zeta_L\, k z_{IR}\,m_{Z_n}^2f^{L3}_n(z)=0,$}\\
   \hbox{$\partial_z(f^{L\pm}_n(z))-\zeta_L\, k z_{IR}\, m_{W_n}^2 f^{L\pm}_n(z)=0,$} \\
    \hbox{$\partial_z(f^{B}_n(z))-\zeta_Y\,k z_{IR}\, m_{Z_n}^2f^{B}_n(z)=0,$}\label{bcz1}
     \end{array}\right.
     \eea
where $\Box Z_\mu^{(n)}(x) = m_{Z_n}^2 Z_\mu^{(n)}(x)$ and
$\Box W_\mu^{(n)\pm}(x) = m_{W_n}^2 W_\mu^{(n)\pm}(x)$.
The equation of motion for the gauge field 5D profiles $f_n(z)$ is
\beq
  \left(\partial_z^2-\frac{1}{z}\partial_z+m_n^2\right)f_n=0.
  \label{bulkeom}
 \eeq
The general solution for the massless zero mode ($m_0=0$) is given by
\beq
 f_0(z)=N_0 + b_0 z^2,
 \label{zerosol}
\eeq
while for the massive mode the general solution is
\beq
     f_n(z)=N_n z (J_1(m_nz)+b_n Y_1(m_nz)),
\label{bulksol}
\eeq
where the coefficients $N_n,b_n$ are fixed by the boundary conditions and normalization condition.

We first check whether the zero mode solution (\ref{zerosol}) satisfies the boundary conditions
(\ref{bcz0}) and (\ref{bcz1}). We find that no zero mode exists for $f_0^{L\pm}$, while for $f_0^{L3},f_0^B$
there is a constant zero mode solution
\beq
    f^{L3}_0(z)=\frac{N_0}{g_{L5}}; \qquad  f^B_0(z)=\frac{N_0}{g_{Y5}},
\label{photon}
\eeq
where $N_0$ is fixed by the normalization condition. The photon wavefunction is
$f_\gamma(z) = \sqrt{(f_0^{L3})^2+ (f_0^{B})^2}$.

Next we check the massive mode solutions, $f^{L\pm}_n(z)$, $f^{L3}_n(z)$ and $f^B_n(z)$.
The boundary conditions determine $m_n$ and $b_n$, while the
overall prefactors $N_n$ are fixed by the normalization condition (which we will do later).
To simplify the expressions, it is also useful to define new variables:
$\beta_5\equiv g_{L5}/g_{Y5}, x_n\equiv m_nz_{IR},t\equiv z_{UV}/z_{IR}$.
Due to the nontrivial boundary conditions and Bessel function properties,
numerical techniques are normally required to solve fo $m_n$ and $b_n$.
However, since we expect there are ultra-light lowest-lying KK modes
satisfying $m_1z_{UV}\ll m_1z_{IR}\ll1$, corresponding to the $W/Z$-bosons,
we can use a small argument expansion for the Bessel functions to obtain a
good analytic approximation for the $W/Z$-boson masses. This also provides
an extra check for the existence of light lowest-lying KK modes.

Consider first the analytic solution for the $W$-boson tower wavefunctions ($f^{L\pm}_n$).
The coefficients $b_n^{L\pm}$ are given by
\beq
 b_n^{L\pm}=-\frac{J_1(x_n t)}{Y_1(x_n t)},
\eeq
and the KK masses are the zeroes of the algebraic equation
\beq
   Y_1(x_n t)[J_0(x_n)-k\zeta_Lx_nJ_1(x_n)]- J_1(x_n t)[Y_0(x_n)-k\zeta_Lx_nY_1(x_n)]=0.
\eeq
The $W$-boson is identified with the lowest-lying KK mode and an approximate expression is given by
\beq
   m_W\simeq\sqrt{\frac{2}{\zeta_L k}}\, z_{IR}^{-1}.
\label{wmass}
\eeq
Therefore to obtain the observed value $m_W=80.4$ GeV, assuming $z_{IR}^{-1}=1$\,TeV,
we need $\zeta_Lk\simeq 310$. Using $b_1^{L\pm}\simeq (\pi t^2)/(2k\zeta_L)$ the $W$-boson
5D profile to leading order becomes
\beq
 {\tilde f}_W(z)=\frac{1}{2}m_Wz^2.
 \label{wprofile}
\eeq
where $f_W(z)\equiv f_1^{L\pm}(z)$ and we write $f_W(z)\equiv N_W {\tilde f}_W(z)$ with $N_W$ a normalization constant.
Thus, with respect to a flat metric the profile is peaked towards the IR brane.

\begin{figure}
\begin{center}
        \includegraphics[width = 0.7\textwidth]{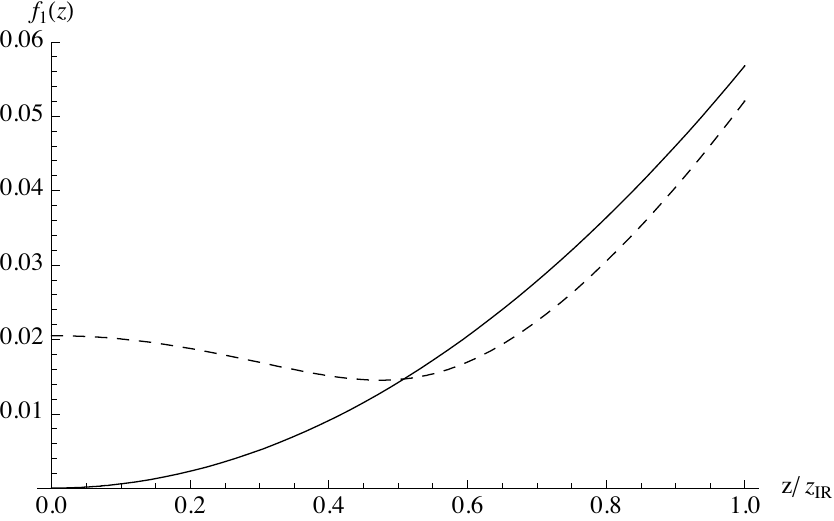}
\end{center}
\caption{The $W$-boson (solid) and $Z$-boson (dashed) profiles in units of $\sqrt{k}$.}
\label{WZfig}
\end{figure}

Due to the nontrivial boundary conditions, the solution for the $Z$-boson tower is not as simple.
The expressions for the wavefunction coefficients are
\bea
 b_n^{L3,B}&=&\frac{\zeta_{L,Y} k x_n J_1(x_n)-J_0(x_n)}{Y_0(x_n)-\zeta_{L,Y} k x_nY_1(x_n)}
 \stackrel{n=1}\simeq\frac{(\frac{1}{2}\zeta_{L,Y}kx_1^2-1)\frac{\pi}{2}}{\log(\frac{x_1}{2})+\gamma_E+\zeta_{L,Y}k}, \label{coeff}\\
 \frac{N_n^B}{N_n^{L3}}&=&\beta_5\frac{J_1(x_n t)+b_n^{L3}Y_1(x_n t)}{J_1(x_n t)+b_n^{B}Y_1(x_n t)}
  \stackrel{n=1}\simeq\beta_5\frac{b_n^{L3}}{b_n^B},
\label{coeffalp}
\eea
where $\gamma_E\simeq 0.577$ is the Euler-Macheroni constant.
Substituting the 5D profiles into the first line of the boundary condition (\ref{bcz0}), gives the
$Z$-boson mass equation:
\beq J_0(x_n t)+b_n^{L3}Y_0(x_n t)+\beta_5
\frac{N_n^B}{N_n^{L3}}(J_0(x_n t)+b_n^{B}Y_0(x_n t)) +\zeta_Q
k(1+\beta_5^2)x_n t(J_1(x_n t)+b_n^{L3}Y_1(x_n t))=0. \label{zeqn}
\eeq
Using the expressions (\ref{coeff}) and (\ref{coeffalp}), an
approximate solution of (\ref{zeqn}) can be obtained in the limit of
vanishing $\zeta_Y$ and large $\zeta_Q$ (i.e. $\zeta_Q\gg\log t$),
with $\beta_5\sim {\cal O}(1)$ (as we will see this limit also helps
to obtain a good fit to the $T$ parameter). This leads to the
$Z$-boson mass:
\beq
    m_Z\simeq \sqrt{\frac{2}{\zeta_Lk}+\frac{2}{\zeta_Qk(1+\beta_5^2)}}\,z_{IR}^{-1}~.
\label{zmass}
\eeq
Assuming $z_{IR}^{-1}=1$~TeV, the observed $Z$-boson mass, $m_Z=91.2$ GeV,
is obtained for $\zeta_Qk\simeq 500$. Note that in the large $\zeta_Q$ or large
$\beta_5$ limit, $m_Z=m_W$, suggesting that there is a custodial symmetry in our model,
as we will show later. An approximate analytic expression for the 5D $Z$-boson profile can
also be obtained in the above limit from $f_1^{L3}$ and $f_1^B$, giving rise to
\bea
 {\tilde f}^{L3}_1(z)&=&\frac{1}{2}m_Zz^2-\frac{m_Z^{-1}}{\zeta_Qk(1+\beta_5^2)},\label{zprofile0}\\
 {\tilde f}^B_1(z)&=&-\frac{\beta_5\log(m_Zz_{IR})}{2\zeta_Qk(1+\beta_5^2)}m_Zz^2-\frac{\beta_5 m_Z^{-1}}
 {\zeta_Qk(1+\beta_5^2)},
 \label{zprofile}
\eea
where $f_1^{L3,B}\equiv N_Z {\tilde f}_1^{L3,B}$ and $N_Z$ is a normalization constant.
The $Z$-boson profile is then given by $f_Z(z) = \sqrt{ (f_1^{L3})^2 +(f_1^B)^2}$ and the
$W/Z$-boson profiles are plotted in Figure~\ref{WZfig}. Thus we again see that with
respect to a flat metric, just like the $W$-boson, the $Z$-boson is localized near the IR brane.

It is important to note that the bulk profiles (\ref{wprofile}), (\ref{zprofile0}),
and (\ref{zprofile}), plotted in Figure~\ref{WZfig} do not depict the complete
localization information. Recall that the profile density $|f(z)|^2$ represents
the probability of finding a particular mode in the AdS slice (including the branes).
The canonical normalization of the profile ensures a unit probability
of locating the mode somewhere in the AdS slice. However the bulk
profiles depicted in Figure~\ref{WZfig} do not give rise to a unit probability
distribution, in fact they only contribute a small fraction to the whole normalization.
This is because the contributions from the boundary kinetic terms are not shown.
They actually give rise to a large contribution to the normalization
which together with the bulk contribution leads to a unit probability distribution.

Since the boundary kinetic terms are given in terms of a singular
$\delta$-function their contribution cannot be depicted in
Figure~\ref{WZfig}. This is related to the assumption that the
branes are infinitely thin, which is just a mathematical
approximation in the low energy EFT. However we know that physically
branes must have a finite thickness, related to the 5D cutoff scale $M_5$,
where the brane-generating dynamics becomes important. The origin of
the 3-branes can be field theory domain walls~\cite{Sundrum:1998sj,
Carena:2002me,delAguila:2003bh}, or effectively arising from warped
throats in string theory~\cite{Giddings:2001yu}. Consequently, assuming
a finite brane thickness, the $\delta$-function localized boundary kinetic term
should actually be represented by a smoothed-out profile within the brane
thickness. For the $W,Z$ profiles this gives rise to a much sharper
but finite peak around the IR brane than is naively depicted in
Figure~\ref{WZfig}. Even though the profile within the brane
thickness cannot be determined in our EFT approach, we know that the
integral over the profile density within the IR brane should match
$\zeta (f(z_{IR}))^2$, where $f(z_{IR})$ is the bulk solution
evaluated at the IR brane. As we will see later, the brane thickness
has a nontrivial influence for matching to the Standard Model.

Although we have focused on the ultra-light first KK modes, which
are identified with the $W/Z$ bosons, it is necessary to check
whether the higher KK modes are compatible with experimental
constraints. Numerically we find that the second KK $W/Z$-boson is
typically around 4 TeV. However if $\zeta_Y$ is large, the first KK
photon (recall that the photon KK tower has a massless zero mode)
can be much lighter than the TeV scale, since as we have seen, a
large brane kinetic term is responsible for an ultra-light KK mode.
Such a light KK photon conflicts with the known bounds on the $Z'$
boson mass~\cite{Amsler:2008zzb}, although it can be partly
compensated by the reduced coupling arising from the brane kinetic
terms which suppress the KK wavefunctions on the
boundary~\cite{Carena:2002dz, Davoudiasl:2002ua}.
Nevertheless to be safely within experimental bounds we choose
small $\zeta_Y\sim 0.1k^{-1}$ so that the first photon KK state has
a mass around 2 TeV. Its coupling to fermions is similar to that of
the zero-mode photon although it depends on the details of the
fermion localization. In the next section, we will see that a small
$\zeta_Y$ is also necessary to ensure a custodial protection for the
$T$ parameter of the model.

Therefore we see that to obtain the correct $W,Z$-boson masses while
maintaining compatibility with other constraints the brane kinetic term
coefficients cannot all be of the same order, and there must be a
small hierarchy between $\zeta_Y$ and $\zeta_L,\zeta_Q$, of
approximately $10^{-3}$. However as argued earlier, brane kinetic
term coefficients are free parameters determined by unknown strong
dynamics, and all the chosen values are `reasonable' from an effective field
theory point of view, so there should be no problem with the small hierarchy
between $\zeta_Y$ and $\zeta_L,\zeta_Q$.

\subsubsection{Fermions}
As mentioned earlier, we are focusing on composite weak gauge bosons
in this paper. Nevertheless it behooves us to comment on the
fermions since they are an integral part of a complete mass generation
framework for the SM. If fermions are to obtain their masses from
EWSB on the UV brane then they should have a nonzero coupling there.
The simplest solution would be to confine them on the UV brane.
However this is not a realistic solution because it would
lead to vanishing gauge couplings since the $W,Z$-bosons have
Dirichlet boundary conditions there. Instead the fermions must be
bulk fields with the same profile for all flavors. This is because
the non-flat electroweak gauge boson profiles (see
Fig.~\ref{WZfig}) require the fermion profiles to be universal in
order to ensure a universal gauge coupling. Unlike models with flat
gauge-boson profiles where gauge coupling universality is automatic
once fermion profiles are canonically normalized, a non-flat gauge
boson profile causes the gauge-boson-fermion-fermion overlap
integral (corresponding to the 4D gauge coupling) to be very different
for different fermion profiles. Moreover requiring a universal fermion profile no
longer allows the fermion mass hierarchy to be explained by the
``geography" in the warped dimension~\cite{Grossman:1999ra,
Gherghetta:2000qt}.

There is an alternative possibility which is to introduce fermion
mass terms on the UV brane. With Planck-scale boundary masses
and IR localization a common but $\cal O$(TeV) mass scale can be obtained
for the zero mode fermions. To distinguish between the flavors,
a new flavor symmetry that is broken on the UV brane can be introduced.
The fermion mass hierarchy can then be explained by a Froggatt-Nielsen
mechanism on the UV brane with different charge assignments when
the flavor symmetry is broken. Assuming a universal bulk fermion profile 
(ensured with a universal bulk mass term), approximate universal gauge 
couplings are then preserved as long as the fermions are light compared 
to the IR scale, $z_{IR}^{-1}$. Interestingly, the slight deviations from
universality of the heaviest fermions could have a palliative effect on
small strains among the $Z\to b\bar b$ observables. We postpone a
detailed study of the fermion sector since it is outside the scope of the 
present work. But for concreteness we assume that the fermions are 
confined to the IR brane in order to perform an electroweak precision 
analysis of the gauge-boson sector. This approximates a bulk fermion profile
that is peaked near the IR brane.

\subsection{EWSB and the dual 4D interpretation}

In order to facilitate a deeper understanding of the 5D model we
discuss in more detail how electroweak  symmetry is broken and
present the 4D dual interpretation via the AdS/CFT correspondence. This
will also help to understand some aspects of precision electroweak
tests in the next section. The underlying physics of our 5D model is
different from other existing models of EWSB in warped space. In the
5D model EWSB occurs on the UV brane via Dirichlet boundary
conditions. This leads to $W/Z$-bosons appearing as ultra-light
first KK mode states that arise from introducing brane kinetic
terms, while higher KK modes are of order the TeV scale, consistent
with experimental constraints and can possibly be seen at the LHC.

It is interesting to compare our model with the technicolor-like  5D
Higgsless model~\cite{Csaki:2003zu}. In the Higgsless model EWSB
occurs on the IR brane via Dirichlet boundary conditions, and the
$W/Z$-bosons are identified with the first KK modes that are light
due to a suppression from the logarithmn of the warp factor. At
first glance it appears that our model is just the Higgsless model
where the UV and IR boundary conditions have been interchanged,
together with the addition of brane kinetic terms. However the
underlying physics is quite different, which is made clearer by
describing the model in the 4D dual picture. As noted in
\cite{Csaki:2003zu}, a Dirichlet boundary condition for a gauge
field is equivalent to a modified Neumann boundary condition where
the gauge field is coupled to a brane Higgs field in the large VEV
limit (which decouples the Higgs boson rendering the low-energy
theory ``Higgsless"). Therefore an intuitive way to see how the
$W/Z$-bosons obtain a mass is to vary the VEV of the brane Higgs
field.

When the 4D electroweak symmetry is restored by a zero VEV in the
technicolor-like 5D Higgsless model, the bulk
gauge fields have exact massless zero modes that are identified with
the $W/Z$-bosons. The higher $W/Z$ KK modes are massive and
degenerate, obtaining their mass from the breakdown of conformal
symmetry (due to the presence of the IR brane). When the VEV is
switched on and the electroweak symmetry is broken, the mass
spectrum gets deformed. The original massless zero modes obtain an
electroweak scale boundary mass that arises entirely from the gauge
coupling to the IR Higgs. In contrast, the higher KK masses are only
slightly shifted (not `generated') due to the coupling to the IR
Higgs, with the major portion of their mass still due to the
breaking of conformal symmetry. In the Higgsless limit, only the
Higgs field itself decouples, while the original gauge field zero
modes remain in the spectrum as the lowest-lying KK state (which are
identified as the massive $W/Z$-bosons via the usual Higgs mechanism).

Thus, the point we would like to stress is that in the Higgsless model, and
other existing EWSB models in warped space,
the contribution to the $W/Z$-boson mass is entirely from the boundary Higgs mechanism.
Alternatively, in the 4D dual description, the $W/Z$-bosons are elementary gauge
fields\footnote{It is straightforward to check that in the technicolor-like Higgsless model,
although the massive $W/Z$-bosons are the lowest-lying KK modes, their wavefunction is localized
towards the UV brane, in contrast to the higher KK modes which are peaked on the IR brane.
In this sense it is dual to a mostly elementary 4D field.},
that are `external' to the strongly coupled CFT. The breaking of conformal symmetry at the
TeV scale triggers EWSB and generates the $W/Z$-boson masses through the electroweak
gauge interaction between the $W/Z$-bosons and some Higgs-like field (or in technicolor
language, the techni-pions). All of the mass is due to electroweak symmetry breaking via a
Higgs mechanism.

By contrast in our model EWSB occurs  by imposing Dirichlet boundary conditions on the
UV brane. Again these boundary conditions can be interpreted as modified Neumann conditions
with a UV brane Higgs and Planck scale VEV. To see the role played by a UV Higgs we can
compare the mass spectrum with that obtained when electroweak symmetry is restored.
For simplicity consider just the $W$-boson and change the UV Dirichlet boundary
conditions in (\ref{bcuv}) to be pure Neumann, while still allowing nonzero brane kinetic terms.
As expected the mass spectrum now contains a massless mode and a light first KK mode due to
the brane kinetic term with mass $m_1\simeq \sqrt{2/(\zeta_L k) + 2/(\pi k R)} z_{IR}^{-1}\sim 3\,m_W$,
where $m_W$ is the $W$-boson mass given in (\ref{wmass}), while the higher KK modes masses just
get shifted at the $1\%$ level.

To study how the mass spectrum in the electroweak symmetric limit changes we consider
a UV brane Higgs with a VEV. The mass spectrum will change depending on how the KK
modes couple to the UV brane. Let the profile of the massless mode and first KK mode be
denoted by $f_0^W(z),f_1^W(z)$, respectively. The $f_0^W$ profile is constant, while the profile
$f_1^W$ peaks towards the UV brane with boundary value $f_1^W(z_{UV})\sim 3 f_0^W(z_{UV})$.
The localization of the first KK mode near the UV brane is due to the IR brane kinetic term.
The remaining KK modes are peaked towards the IR brane. Since only the massless zero mode
and first KK mode largely overlap with the UV brane it is a reasonable approximation to consider
this two-state system coupling to the UV brane Higgs.
Suppose that the UV Higgs boson has a VEV $v$, then the $2\times2$ mass-squared matrix is
\beq
    \mathcal{M}^2\simeq\left(
         \begin{array}{cc}
           0+g_{L5}^2 v^2 f_0^2(z_{UV}) & g_{L5}^2 v^2 f_0(z_{UV})f_1(z_{UV})\\
            g_{L5}^2 v^2 f_0(z_{UV})f_1(z_{UV}) & m_1^2+g_{L5}^2 v^2 f_1^2(z_{UV})\\
         \end{array}
       \right)
       =\left(
         \begin{array}{cc}
           \epsilon_0^2 {\hat v}^2 & \epsilon_0\epsilon_1 {\hat v}^2 \\
            \epsilon_0\epsilon_1 {\hat v}^2& m_1^2+\epsilon_1^2 {\hat v}^2  \\
         \end{array}
       \right),
\eeq
where $f_0(z_{UV})= \epsilon_0 \sqrt{k}$ with $\epsilon_0\simeq 1/\sqrt{\zeta_L k+\pi k R}$, $f_1(z_{UV})=\epsilon_1\sqrt{k}$ with $\epsilon_1 \simeq 1/\sqrt{\pi k R}$ and ${\hat v}^2=g_{L5}^2 k\, v^2$. After 
diagonalization, in the limit $v\gg m_1$ and $m_1/v \ll \epsilon_{0,1}$, the two mass eigenvalues are
\bea
       \widetilde{m}_0^2&=& (\epsilon_0^2 +\epsilon_1^2) {\hat v}^2+{\cal O}(m_1^2)\simeq
       \frac{g_{L5}^2}{\pi R} v^2,\label{lightzero}\\
       \widetilde{m}_1^2&=&\frac{\epsilon_0^2}{\epsilon_0^2 +\epsilon_1^2} m_1^2+{\cal O}\left(\frac{m_1^4}{v^2}\right)
       \simeq \frac{\pi R}{\zeta_L} m_1^2.
\label{lightestwo}
\eea
We see that in the limit $v\rightarrow \infty$, where the UV Dirichlet boundary conditions are restored, there is one heavy mode that decouples from the low-energy 4D theory, while a light KK mode remains in the spectrum (which is identified as the $W$-boson). Note that in (\ref{lightzero}) the heavy mode $\tilde{m}_0$ obtains a mass proportional to $v$ via the usual Higgs mechanism associated with the UV Higgs. It is the counterpart of the $W/Z$-boson in the 5D Higgsless model -- they both originate from the zero mode before EWSB, although strictly speaking it is not precise to correlate the new massive mode with just the original zero mode or the first KK mode since both original states are highly mixed due to the UV Higgs. Nevertheless the main difference
is that the original zero modes in our model eventually decouple and become irrelevant to the
low-energy SM since they receive a Planck scale mass from the UV Higgs, while the original zero
modes in the 5D Higgsless model~\cite{Csaki:2003zu} remain in the low-energy theory as the $W/Z$-bosons since they obtain an electroweak scale mass from an IR Higgs.

Instead in our model the $W/Z$-bosons are identified with the light mode whose mass $\tilde{m}_1$
originates from the previous first KK mode. In fact from (\ref{lightestwo}), $\tilde{m}_1$ depends on $m_1$ which is determined by the IR scale and the brane kinetic term coefficients. The actual contribution from the UV Higgs is sub-leading and highly suppressed. So the major difference is that unlike most existing EWSB
models, the $W/Z$-bosons in our model do not obtain a mass from the Higgs mechanism, which breaks electroweak symmetry on the boundary. Instead, except for a reduced mass due to the brane kinetic terms, they are just like
the usual KK modes whose mass originates from the CFT breaking scale regardless of the existence of the
EWSB Higgs at the boundaries. Thus the novel feature of our model is that the $W/Z$-bosons obtain their mass from two different 5D locations: a dominant contribution from conformal symmetry breaking at the IR brane and a sub-dominant contribution from EWSB on the UV brane. The actual mass difference between the $W$ and $Z$-boson arises from the mixing on the UV brane introduced by finite values of $g_{Y5}$ and $\zeta_Q$.

Our model can also be understood from the dual 4D interpretation. Using the AdS/CFT
dictionary~\cite{adscftdict}, the bulk 5D gauge symmetry corresponds to a global symmetry in the 4D CFT.
The unbroken gauge symmetry on the UV brane weakly gauges that particular subgroup of the global symmetry.
Therefore the 4D dual of our model is a strongly-coupled CFT with an $SU(2)_L\times U(1)_Y$ global symmetry,
whose $U(1)_Q$ subgroup is weakly gauged. This suggests that electroweak symmetry is {\it not a fundamental
gauge symmetry}. The $W/Z$-bosons are CFT bound states created by the global current associated with the electroweak global symmetry. The dominant contribution to their masses arises from the IR conformal symmetry breaking scale, while their mass difference results from EWSB in an elementary sector at the UV scale.
To explain their universal coupling to matter as experimentally observed,
we can interpret them as gauge fields of some hidden local symmetry by promoting the previous global
electroweak symmetry to be local, as was similarly done for the $\rho$ meson in QCD -- this is just the usual
Standard Model Yang-Mills theory, which has been proved to be a very successful effective low-energy
description.

If we want to further explore the strong dynamics which `possibly' underlies the SM or its 5D dual, we may ask:
what kind of `special dynamics' gives rise to large brane kinetic terms or their 4D dual counterpart?
Why does this introduce an additional light collective mode? As mentioned in the introduction, in 5D
models with orbifold compactification or bulk fields interacting with brane fields, brane kinetic terms necessarily
emerge as counterterms for loop corrections of the gauge field propagator. Also as demonstrated
in~\cite{Ponton:2001hq}, there is no problem with having a larger brane kinetic term compared to NDA
as needed in our model. In fact the existence of large brane kinetic terms is consistent with our model
where a large number of matter fields are localized on the IR brane assuming that brane kinetic terms
originate from interactions with brane localized matter \cite{Dvali:2000rx}. Similarly, on the UV brane a
large brane kinetic term could result from integrating out a large number of string states. The 4D dual
description of brane kinetic terms are bare kinetic terms emerging at the appropriate cutoff scales.
According to the AdS/CFT correspondence, the 4D dual description
of obtaining an IR brane kinetic term from a one-loop counter-term is
related to sub-leading large-$N$ effects for the corresponding operator
correlation function in the CFT \cite{Agashe:2002jx}. Of course, there could be other possibilities
leading to large brane kinetic terms, including large volume effects at the brane locations, or
possibilities related to the physics of stabilizing the radion~\cite{Ponton:2001hq}.
In Appendix A we use the 4D KK modes to demonstrate more clearly how an
ultra-light mode generically arises with a large brane kinetic term compared
to directly solving the 5D equations of motion. By introducing a brane kinetic term,
the 4D KK spectrum changes due to kinetic mixing and mass mixing induced by the boundary
kinetic term.

At this point it is instructive to comment on the relation of our model to the original
RS1 model with the Standard Model confined on the IR brane~\cite{Randall:1999ee}.
We have seen that for fixed $z_{IR}$, specific values of $\zeta_L,\zeta_Q$
can give rise to realistic non-zero $W,Z$-boson masses. It is interesting to study
the limit of large brane kinetic terms (with $z_{IR}$ fixed). The $W,Z$-bosons in our
model then become increasingly confined to the IR brane. They are also becoming lighter,
while the remaining Kaluza-Klein modes are becoming increasingly heavier. Eventually
as the brane kinetic terms become infinite the remaining Kaluza-Klein modes decouple
and we are left with massless $W,Z$ bosons confined on the IR brane. This can be seen
in more detail by considering, for example, the $W$-boson. Naively we would expect the limit
$\zeta_L\rightarrow\infty$ to be singular. However, from Eq.(\ref{5daction}) the
normalization of the $W$-boson profile is given by
\beq
     N_W^2\left[\int \frac{dz}{kz}({\tilde f}_W(z))^2+\zeta_L({\tilde f}_W(z_{IR}))^2\right]=1.
\label{normW}
\eeq
Substituting the expressions for the $W$-boson mass (\ref{wmass}) and profile
(\ref{wprofile}), we find that when $\zeta_L\rightarrow\infty$, the bulk integral part of
(\ref{normW}) vanishes, while the boundary kinetic term part is finite and can be normalized to one.
This means that a massless $W$-boson is completely localized on the IR brane.
This is similar to RS1 when the gauge bosons are massless, except that our photon
is an elementary field with a large localization on the UV brane whereas in RS1 the
photon is also confined to the IR brane.
Of course the major difference between our model and RS1 is the way in which the
$W,Z$-bosons obtain a mass. In RS1 there is an associated Higgs sector also
confined on the IR brane to give mass to the electroweak gauge bosons.
By contrast in our model the $W,Z$-bosons obtain a mass from finite boundary
kinetic terms that allows the $W,Z$-bosons to obtain a bulk profile with a corresponding
KK mass. Therefore we see that brane kinetic terms generalize the usual setup,
providing a way to interpolate between the localized bulk profile and the boundary fields.

Finally we briefly comment on the interesting possibility that our scenario admits
an `effective' dual description in terms of a Higgs mechanism
implemented with a non-linear $\sigma-$model, where $\zeta_L^{-1},
\zeta_Q^{-1}$ play a role similar to the VEV (see (\ref{wmass}) and (\ref{zmass})).
This is analogous to the dual description that can be made when interpreting the
QCD $\rho$ meson as a massive gauge boson of spontaneously broken
hidden local symmetry~\cite{Bando:1984ej}. In more `modern' language,
this suggests that our composite 4D model could be a non-supersymmetric example
of Seiberg duality~\cite{Seiberg:1994pq, Harada:2003jx}, where the underlying
strong confining gauge theory has a low-energy dual description as
a weakly-coupled gauge theory in the Higgs phase, and the emergent
composites are effective degrees of freedom. However to understand this better
requires a detailed knowledge of the constituent gauge theory.

\section{Electroweak Precision Analysis}

Just like any new physics beyond the SM, our model needs to be
consistent with a precision electroweak analysis. As in \cite{Cacciapaglia:2004jz}
we focus on oblique corrections, characterized by the $S$ and $T$
parameters, and briefly discuss the $V$ parameter. Although both a light
Higgs boson theory like the SM and technicolor-like Higgsless EWSB models
are well-motivated theoretically, the Higgsless models are disfavored
compared to models with a light Higgs boson. The major reason being that the
$S$ parameter in the technicolor-like Higgsless models is typically large and
positive, which is ruled out by precision electroweak measurements.
Instead, we find that our emergent model can give a reasonable fit to the
$S$ parameter. Furthermore, there is also a built-in custodial symmetry,
so that the $T$ parameter is compatible with experimental constraints.
The better agreement of our model with electroweak precision tests,
especially the $S$ parameter, compared to the usual
Higgsless models originates from their essential differences in the
underlying physics discussed in the previous section.

To calculate the $S$ and $T$ parameters we will use the same matching
scheme as in \cite{Cacciapaglia:2004jz}. For simplicity the fermions are
assumed to be localized on the IR brane where they obtain a nonzero
coupling with the gauge bosons. The radiative corrections will be oblique
by requiring that the couplings of the IR-localized fermions give exactly
the leading order coupling relations. When there are no boundary kinetic
terms the matching of 4D couplings with 5D couplings is simply given by
$g_4=g_5 f_A(z_{IR})$ where $f_A(z)$ is the bulk gauge field profile.
This is the case that has been considered in the literature.
However with boundary kinetic terms there is a  $\delta$-function contribution
to the profile as mentioned in Section 2.1.1. Its contribution can be incorporated
into the matching by including the physical brane thickness $\Delta \simeq (0.01-0.1)
z_{IR}^{-1}$, as mentioned earlier. The matching now becomes
\beq
   g_4=g_5\int^{z_{IR}}_{z_{IR}-\Delta} \frac{dz}{(kz)^4}\, f_\psi(z)f_\psi(z)\bar{f}_A(z),
\eeq
where $f_\psi$ is the fermion profile, and $\bar{f}_A$ is the modified IR
profile of the gauge field satisfying
\beq
   \int^{z_{IR}}_{z_{IR}-\Delta}\frac{dz}{kz}\,(\bar{f}_A(z))^2=\zeta (f_A(z_{IR}))^2,
\eeq
for a brane kinetic term coefficient $\zeta$. This effectively means that the bulk
gauge field profile at the IR brane, $f_A(z_{IR})$, is scaled by a factor $\sqrt{1+\delta}$,
where $\delta\simeq 10-100$ depends on the specific profile within the brane thickness and
the brane kinetic term coefficient. In our EFT approach it is an undetermined parameter
that regulates the underlying dynamics associated with the $\delta$-function singularity
of the IR brane. In order to respect the bulk and brane symmetries we find
that this regulator needs to be universal for all gauge field profiles.

The couplings of an $SU(2)_L$ doublet fermion to gauge bosons can
then be obtained from the bulk covariant derivative on the IR brane.
\bea
&&\left(g_{L5}T_3A_\mu^{L3}+g_{L5}T_\pm
   A_\mu^{L\mp}+\frac{Y}{2}g_{Y5}B_\mu\right)\Bigg |_{z=z_{IR}}\nonumber\\
  &=&\sqrt{1+\delta}\left\{N_0Q\gamma_\mu+g_{L5}f_1^{L\mp}(z_{IR})T_\pm
   W^\mp_\mu+g_{L5}f_1^{L3}(z_{IR})\left(T_3+\frac{g_{Y5}f_1^B(z_{IR})}{g_{L5}f_1^{L3}(z_{IR})}
   \frac{Y}{2}\right)Z_\mu\right\},\quad
   \label{covderiv}
\eea
where we have used the KK decompositions (\ref{KKdec1})-(\ref{KKdec3}) and the photon profile (\ref{photon}).
Note that in (\ref{covderiv}) the scaling factor is mostly effective for the photon and $W,Z$-bosons, since the
normalization of the higher KK modes is primarily from the bulk integral and not the boundary kinetic terms.
The SM fermion hypercharge is denoted by $Y$, the electric charge by $Q=T_3+Y/2$ and $T_\pm$ denote
the weak isospin charge. Note that the universal nature of the scaling factor $\sqrt{1+\delta}$ is crucial in order
to obtain the correct electric charge. To match to the SM, we require that (\ref{covderiv}) reproduce the SM fermion
couplings in terms of the 4D gauge couplings $g,g'$. We first write down two relations that determines the matching
between the 5D and 4D gauge couplings which are independent of the wavefunction normalizations:
\bea
   \frac{g'^2}{g^2}&=&-\frac{g_{Y5}f_1^B(z_{IR})}{g_{L5}f_1^{L3}(z_{IR})}\equiv-\frac{f_1^B(z_{IR})}
   {\beta_5 f_1^{L3}(z_{IR})},\label{gcoupling1}\\
   \frac{1}{e^2}&=&\frac{1}{g^2}+\frac{1}{g'^2}\equiv\frac{1}{(1+\delta)N_0^2}.
\label{gcoupling2}
\eea
The normalization $N_0$ can be easily fixed using the fact that $U(1)_Q$ is unbroken, so the photon kinetic term, with kinetic term coefficient $Z_\gamma$, should always be canonically normalized:
\bea
  Z_\gamma&=&\int^{z_{IR}}_{z_{UV}}\frac{dz}{kz}\left[(f_0^B(z))^2+(f_0^{L3}(z))^2\right]
  +\frac{\zeta_Q}{g_{Y5}^2+g_{L5}^2}\left[g_{Y5}f_0^{L3}(z_{UV})+g_{L5}f_0^{B}(z_{UV})\right]^2\nonumber\\
  &&\qquad +\zeta_L(f_0^{L3}(z_{IR}))^2+\zeta_Y(f_0^B(z_{IR}))^2,\nonumber\\
  &=&\frac{N_0^2}{g_{L5}^2 k}\left[(1+\beta_5^2)\log\left(\frac{z_{IR}}{z_{UV}}\right)
  +\zeta_Q k(1+\beta_5^2)+\zeta_L k+\zeta_Yk\beta_5^2\right]=1.
\label{photonnorm}
\eea
Therefore the full matching between the 4D and 5D gauge couplings is determined
by (\ref{gcoupling1}), (\ref{gcoupling2}) and (\ref{photonnorm}) to be:
 \beq
     g^2=g_{L5}^2k(1+\delta)\left(1-\frac{\beta_5 f^{L3}_1(z_{IR})}{f^B_1(z_{IR})}\right)\left[(1+\beta_5^2)
     \log\left(\frac{z_{IR}}{z_{UV}}\right)+\zeta_Qk(1+\beta_5^2)+\zeta_Lk+\zeta_Yk\beta_5^2 \right]^{-1},
\label{gmatching0}
\eeq
\beq
     g'^2=-g^2\frac{f^{B}_1(z_{IR})}{\beta_5 f^{L3}_1(z_{IR})}.
\label{gmatching}
\eeq
These matching relations can be used to fix the normalization factors for $f^{B}_1(z),f^{L3}_1(z)$
together with the following relations obtained by matching (\ref{covderiv}) to the SM result:
\bea
g_{L5} \sqrt{1+\delta} f_1^{L\pm}(z_{IR})&=&g,\label{fixnorm0} \\
  g_{L5}\sqrt{1+\delta}f_1^{L3}(z_{IR})&=&g\cos\theta_w\equiv\frac{g^2}{\sqrt{g^2+g'^2}},\label{fixnorm}
\eea
where $\theta_w$ is the weak mixing angle. Note that in (\ref{fixnorm0}) the universal
scaling factor associated with the brane thickness does not cancel and leads to the
estimate $g_{L5}^2 k \simeq 132/(1+\delta)$ for the parameters used in
Section 2.1.1. Thus for $\delta\sim 10$ the 5D coupling remains perturbative although
clearly without the brane thickness the 5D theory would be strongly-coupled.

Using the definitions (\ref{zprofile0}) and (\ref{zprofile}) the normalization factor $N_Z$ from
(\ref{fixnorm}) is given by:
\bea
     N_Z&=& \frac{1}{\sqrt{1+\delta}}\frac{g}{g_{L5}} \frac{1}{\sqrt{1+g'^2/g^2}}
     \frac{1}{{\tilde f}_1^{L3}(z_{IR})},\nonumber\\
     &=&\sqrt{-\frac{\beta_5 k}{{\tilde f}_1^{L3}(z_{IR}){\tilde f}_1^{B}(z_{IR})}}\left[(1+\beta_5^2)
     \log\left(\frac{z_{IR}}{z_{UV}}\right)+\zeta_Q k(1+\beta_5^2)+\zeta_Lk+
     \zeta_Yk\beta_5^2\right]^{-\frac{1}{2}}.
\label{znorm}
\eea
As expected the normalization $N_Z$ does not depend on the rescaling factor $\sqrt{1+\delta}$.
Having set up the matching and normalization for our model, we are now ready to begin the
precision electroweak analysis.

\subsection{Custodial symmetry and the $T$-parameter}

As is well known, a major criteria for realistic EWSB model-building is to ensure that the tree-level mass
ratio between the $W$-boson and $Z$-boson satisfies the relation:
\beq
      \rho\equiv\frac{m_W^2}{m_Z^2 \cos^2\theta_w}=1.
\eeq
The deviation from this tree-level prediction due to new physics is
well constrained by precision electroweak data \cite{:2005ema} in
terms of the $T$ parameter, defined by~\cite{Peskin:1991sw}
 \beq
     T\equiv\frac{\rho^*(0)-1}{\alpha}=\frac{4\pi}{\sin^2\theta_w \cos^2\theta_w m_Z^2}
     (\Pi_{11}(0)-\Pi_{33}(0)),
\label{Tdef}
\eeq
where $\rho^*$ is the theory prediction, and $\alpha=1/128.9$ is the fine structure constant
at the $Z$-pole. An automatic way to ensure that $\rho=1$ at leading order and is
well protected from radiative corrections is to introduce a
custodial $SU(2)$ symmetry which ensures that the triplet $A^{L1,2,3}$ masses are degenerate
in the $A^{L3}-B$ decoupling limit when EWSB occurs~\cite{Sikivie:1980hm}.

In the usual models where the $W/Z$-bosons obtain their mass from an
elementary Higgs doublet, a custodial  $SU(2)$ symmetry naturally
appears due to the larger global symmetry of the Higgs potential
$SO(4)\simeq SU(2)_L\times SU(2)_R$, whose diagonal subgroup
$SU(2)_D$ remains unbroken. However in other models like the Little
Higgs model, and the technicolor-like Higgsless model, custodial
$SU(2)$ does not automatically appear, and needs to be introduced to
the simplest scenarios. Similarly, in our model it appears that
there is no custodial symmetry, since the maximal symmetry group is
just the SM electroweak group $SU(2)_L\times U(1)_Y$. However, there
is a special custodial mechanism already built into our simple model
in the small $\zeta_Y$ limit. The custodial symmetry is just the
$SU(2)_L$ itself!

To see how this comes about, consider the limit where $\zeta_Y=0$.
It would appear that when the bulk $SU(2)_L\times U(1)_Y$ symmetry
is broken to $U(1)_Q$, the broken $SU(2)_L$ cannot survive as the
custodial symmetry. However recall that the $W,Z$-bosons do not
originate from the zero modes. Instead they are true KK modes
which obtain a mass from the presence of an IR brane (or the
breaking of conformal symmetry) even in the absence of electroweak
symmetry breaking at the boundaries. At each KK level
there is an $SU(2)$ triplet of massive vector bosons. By contrast in
usual models the zero-mode $SU(2)$ triplet becomes the massive SM triplet
$W^{1,2,3}$ when symmetry-breaking boundary conditions are added.
However without an additional bulk custodial symmetry there is no
guarantee that the $T$ parameter is small for these boundary-generated
masses. Although a SM-like EWSB with custodial symmetry can be
implemented by Higgsing at the IR boundary to ensure $\rho=1$ at
leading order, there is typically a large $\log(z_{IR}/z_{UV})$ enhanced
contribution to the $T$ parameter from the bulk integral
without a bulk custodial symmetry~\cite{Csaki:2002gy, Agashe:2003zs}.
Instead in our case the triplet KK modes, $W^{1,2,3}$ are guaranteed
to be degenerate in mass due to the bulk $SU(2)_L$ symmetry which
acts as a custodial symmetry. This is similar to the isospin symmetry in QCD
which is an approximate symmetry of the $\rho$-meson spectrum.

This is no longer the case if $\zeta_Y$ is nonzero.  The UV mixing between
$A^{L3}$ and $B$ causes them to have a common KK mass $m_{Z_n}$ in Eq.(\ref{bcz1}).
When $\zeta_Y=0$, the IR boundary condition of $f^B$ is independent of $f^{L3}$,
so there is no additional mixing between $A^{L3}$ and $B$. However, with nonzero
$\zeta_Y$, additional non-SM-like mixing occurs between $A^{L3}$ and $B$, parameterized by the same
$m_{Z_n}$. This means that nonzero $\zeta_Y$ causes the $\rho$ parameter to deviate away from one,
and therefore in a realistic model $\zeta_Y$ should stay small relative to $\zeta_Q$ and $\zeta_L$. It cannot be
simply set to zero because no symmetry forbids such a term. Thus, we rely on a small hierarchy between the different brane kinetic term coefficients which can be due to some underlying strong dynamics, and is
consistent from the EFT point of view.

To explicitly check the custodial symmetry and compute the $T$ parameter in our model we use Eq.(\ref{Tdef}).
In 5D models the expressions for the various $\Pi$'s can be obtained directly from the bulk integrals~\cite{Csaki:2005vy}
\bea
g^2\Pi_{11}(0)&=&\int\frac{dz}{kz}\,|\partial_zf_W(z)|^2,\\
(g^2+g'^2)\Pi_{33}(0)&=&\int\frac{dz}{kz}\left[|\partial_zf^{L3}_1(z)|^2+|\partial_zf^{B}_1(z)|^2\right].
\eea
An analytic expression can be obtained but cannot be given in
a simple form with sufficient precision due to the complexity of the expression.
However an analytical fit can be done numerically and
leads to an analytic approximation for the $T$ parameter
\beq
  T\propto \frac{1}{\alpha}(m_Zz_{IR})^2.
\eeq
We can can see that there is no large $\log$ enhancement and depends quadratically on $m_Zz_{IR}$.
Furthermore, the adjustable large brane kinetic term or small hierarchy between $m_W, m_Z$ and
the higher KK scale $\simeq z_{IR}^{-1}$ can give enough suppression for the $T$ parameter to be
compatible with the LEP bound. A benchmark point will be given numerically later which fits both $S$ and $T$.

\subsection{$S$-parameter}

The oblique correction parameter $S$ is defined as
\beq
    S\equiv16\pi(\Pi'_{33}-\Pi'_{3Q}),
\label{s-definition}
\eeq
and is directly related to the wavefunction normalization of $Z$:
\beq
Z_Z=1-\Pi'_{ZZ}=1-(g^2+g'^2)(\Pi'_{33}-2 \sin^2\theta_w\Pi'_{3Q}+\sin^4\theta_w\Pi'_{QQ}),
\eeq
where $\Pi'_{\gamma\gamma}=e^2\Pi'_{QQ}$ and $\Pi'_{\gamma Z}=gg'(\Pi'_{3Q}-\sin^2\theta_w\Pi'_{QQ})$.
Since the photon wavefunction is already canonically normalized, $\Pi'_{QQ}=0$. Furthermore
$\Pi'_{3Q}=0$, since we are doing a tree-level calculation in 5D (corresponding to loop level in 4D),
and there is no $Z-\gamma$ mixing. Thus we obtain:
\beq
    S=\frac{16\pi}{g^2+g'^2} (1-Z_Z),
\label{znorm2}
\eeq
where $Z_Z=1-(g^2+g'^2)\Pi'_{33}$. The wavefunction normalization, $Z_Z$ is calculated by
integrating the bulk gauge profiles, as well as including the boundary kinetic term contributions:
\bea
Z_Z&=&N_Z^2\left\{\int\frac{dz}{kz}\left[({\tilde f}^{L3}_1(z))^2+({\tilde f}^{B}_1(z))^2 \right]+\frac{\zeta_Q}
{1+\beta_5^2}\left[ {\tilde f}^{L3}_1(z_{UV})+\beta_5 {\tilde f}^{B}_1(z_{UV}) \right]^2\nonumber\right.\\
&&\qquad\qquad\left.+\zeta_L({\tilde f}^{L3}_1(z_{IR}))^2+\zeta_Y({\tilde f}^{B}_1(z_{IR}))^2\nonumber \right\},\\
&\simeq&1+ {\cal O}(10) (m_Zz_{IR})^2.
\label{znorm3}
\eea
where the analytic expression is to leading order in $m_Zz_{IR}$. Again the analytical fit
is done by numerical evaluation.
Using (\ref{znorm2}), the analytic expression for the $S$ parameter thus becomes:
\beq
S\propto \frac{16\pi}{g^2+g'^2}(m_Zz_{IR})^2.
\label{s-estimation}
\eeq
This expression reveals several important features of $S$ in our model:
$S$ is always positive and lowering $m_Z z_{IR}$ is the most efficient
way to obtain a small $S$. This can be intuitively understood using the
4D dual interpretation. The $S$ parameter shift arises from the self-energy
diagram where the $Z$-boson mixes with the KK modes below the cutoff scale
which then couples to a fermion loop. With larger KK masses--scaling like
$z_{IR}^{-1}$--this shift is suppressed by $m_{KK}^2$. Therefore, as the mass
difference between the $Z$-boson (which is a special light KK mode) and the
higher KK mode gets larger, the contribution from higher KK modes to $S$ tends
to decouple due to the mass suppression.

It should be emphasized that our model and technicolor-like
Higgsless models face a common challenge to obtain a sufficiently
adequate $S$ parameter. This arises from the sum over KK modes below
the cutoff scale, which gives a factor $N$--the number of KK modes
below the cutoff scale, proportional to the number of colors in the 4D dual theory.
For the AdS/CFT duality to be valid, $N$ has to be large which
implies a large $S$ in general. So additional suppression is always
required. In our model the extra suppression factor comes from the
small hierarchy between $m_Z$ and the IR scale (higher KK mass),
which is realized by introducing brane kinetic terms -- an ingredient
already built into the model. Another possibility is to reduce the
fermion couplings to KK modes by considering fermions with a flat
profile \cite{Cacciapaglia:2004rb}. This approach is not taken in
our EWSB scenario because fermions should have profiles which are
peaked towards the IR brane.

Let us now numerically compare our result for the $S$ parameter with
experiment.  Assuming $z_{IR}^{-1}=1\, {\rm TeV}$, we find that
$S=0.3$, which when compared to the LEP bound \cite{:2005ema},
suggests a higher IR scale is needed to fit precision tests. A
benchmark point which fits the $T-S$ $68\%$ probability contour
according to LEP data \cite{:2005ema} is obtained with the input
parameter set: $z_{IR}^{-1} =1.8\, {\rm TeV}, \zeta_Lk \simeq 1000,
\zeta_Qk\simeq 1700, \zeta_Yk\simeq 0.2$. This gives the correct
$W,Z$-boson masses and $S\simeq 0.1, T\simeq 0.05$. Note that when
comparing with the LEP contour, we subtracted the contribution from
$m_H=114$ GeV, which together with $m_t=178$ GeV defines the
reference point at the origin. This differs from how this
contribution is treated in technicolor-like Higgsless models. In
these models the $W/Z$-bosons obtain a mass from the Higgs mechanism
even though the theory becomes `Higgsless'. This means that an extra
TeV-scale heavy Higgs contribution must be added to the $S,T$ values
which causes a preference for a slightly negative $S$ and positive
$T$ \cite{Cacciapaglia:2004rb}. However, in our model we do not need
to add such a contribution from a heavy Higgs. As discussed in
Section 2, the $W/Z$-boson masses in our model do not arise from the
Higgs mechanism. Instead their mass originates from the IR
scale, or the conformal symmetry breaking scale in the 4D dual
interpretation. There is a usual Higgs mechanism on the UV brane
which gives a UV scale mass to the original zero mode gauge field
causing it to decouple. But the UV Higgs mechansim is not
responsible for generating the $W/Z$-boson masses. Therefore a small
positive $S$ is sufficient to satisfy the precision tests in our
model.

Note that even though a higher IR scale, $z_{IR}^{-1}=1.8\,{\rm TeV}$
was needed to obtain reasonable agreement with precision tests there is a
drawback of increasing the IR scale.
It diminishes the chances of discovering new states such as heavy KK gauge
bosons at the LHC--for example, the next lightest $W$ and $Z$-boson KK mode
masses are increased to $\sim 7$ TeV, although there is still a lighter
KK photon with mass 3.6 TeV, which might be seen at the LHC~\cite{Agashe:2007ki}.
Moreover, note that the IR scale was obtained by fitting the 68\% CL contour.
Using the less restrictive 95\% CL contour can result in a lower IR scale and
therefore increase the chances of detecting KK resonances at the LHC.
The idea of suppressing the $S$-parameter by adding brane kinetic terms and
increasing the KK scale has also been considered in technicolor-like
Higgsless models \cite{Cacciapaglia:2004rb}. However the effect was of
limited use in these models because KK modes were required to be lower
than 1.8 TeV in order to ensure  tree-level unitarity and
calculability. Interestingly, this is not a concern for our model.
As we will demonstrate in the next section, $WW$ scattering
has a very different story in our model: due to the composite nature
of the $W$-boson, tree-level unitarity may
break down earlier than the SM prediction, and we expect an
overall form factor suppression to restore unitarity of $W,Z$ boson
scattering non-perturbatively. We never rely on KK modes
to help maintain perturbative unitarity, so there is no worry about them being heavy.
These issues will be explained in detail in the following section.

Finally we briefly comment on another electroweak precision test--the
$V$ parameter which measures the correction to the Fermi coupling
constant $G_F$ via a four-fermion operator associated with $\mu$ decay:
$G_F=G_{F,W}(1+V)$.
The $V$ parameter can lead to stringent constraints in RS1-type models
with fermions localized on the IR brane because the exchanged KK modes
are universally strongly coupled to the IR brane~\cite{Csaki:2002gy}.
However, in our model we expect a negligible shift of the $V$ parameter
from $W$-boson KK modes. As mentioned earlier
large brane kinetic terms suppress higher KK modes near the IR brane
(recall that as $\zeta\rightarrow\infty$ we obtain an RS1-like scenario where only the
lowest KK mode is confined on the IR brane, while higher modes completely decouple).
To estimate the $V$ parameter consider the next-heaviest KK mode
above the $W$-boson, which is the dominant contribution since higher KK modes are more
decoupled. Numerically we find that for the next-heaviest KK mode
$f_2^{L\pm}(z_{IR})\simeq 10^{-2} f_1^{L\pm}(z_{IR})$. This does not include
an enhancement from boundary kinetic terms to the normalization of the lightest KK
mode at the IR brane, which can be up to ${\cal O}(10)$.
In addition there is a suppression in the propagator
from the relatively large KK mass difference:
$(m_W/m_W^{(2)})^2\simeq 10^{-4}$. Combining all these factors leads
to the rough estimate $V\lesssim10^{-8}$. This is below the upper experimental
constraint but note that our estimate is crude and clearly will depend on the
fermion details. Furthermore neutral current processes which involve KK
photons could lead to more stringent constraints since these KK modes
may not be sufficiently suppressed at the IR brane. Again the details depends on the fermions
and will be postponed for a future analysis.

\section{$WW$ Scattering}
\label{wwsection}

\subsection{Form factors}

One particular type of process within the Standard Model that calls
for new physics at the TeV scale is longitudinal $W$ (or $Z$)-boson
scattering. The tree-level perturbative amplitude $\cal A$ of an individual
graph of such processes involves divergences up to $E^4$, (where $E$
is the center-of-mass energy): ${\cal A}=A(E/m_W)^4+B(E/m_W)^2+C$,
where $A,B,C$ are constants.
In a true gauge theory like the SM where the $W$-boson is an elementary
gauge field, the $E^4$ divergence vanishes due to gauge cancelation
between the contact graph and $s,t$ channel $\gamma,Z$ exchanges. But
the next leading divergence, $E^2$ does not vanish within the SM alone,
and tree-level unitarity breaks down at 1.2 TeV, requiring new
non-perturbative physics to restore unitarity. However, due to the difficulty
with quantitatively describing physics involving strong dynamics, there
has always been a strong preference to preserve tree-level unitarity
or perturbative calculability up to high energy scales. This preference
makes introducing an elementary Higgs boson to the SM a desirable
scenario. In particular, adding graphs involving the Higgs boson
cancels the $E^2$ divergence~\cite{Lee:1977eg}, while the remaining
constant piece can be perturbative with a Higgs boson lighter
than 700 GeV~\cite{Gunion:1989we}. Similarly in alternative EWSB
scenarios like 5D Higgsless models, preserving tree-level unitarity has
been strongly preferred, where summing over KK modes plays a
similar role as a Higgs boson and cannot be too heavy.

However, it should be emphasized that as a foundation of
quantum field theory unitarity itself is never jeopardized by $WW$
scattering concerns. Preserving tree-level unitarity is a
theoretical preference that avoids having to deal with
strongly-coupled theories. It is by no means the choice that has to be
taken by Nature. In fact almost equally importantly, current experimental
constraints on the behavior of $WW$ scattering are rather moderate due
to the energy scale and luminosity reach of colliders. As will be shown
in the next section LEP and Tevatron data only constrains the trilinear
gauge boson coupling to be SM-like up to energies not much beyond the
$WW$ production threshold. So this allows enough room
for theoretical model building and the LHC to test for
possible deviations from the Standard Model predictions at high energy.

Based on these considerations the breakdown of tree-level $WW$
unitarity at lower energy scales is not sufficient to veto a theory,
especially if it can also give quantitative insights into how unitarity is
restored by the strong dynamics. This is the case for our emergent model where
energy-dependent form factors of trilinear and quartic gauge boson
self-interactions are naturally associated with composite $W,Z$ bosons
which can lead to a distinctive explanation of $WW$ scattering and its
unitarization. In particular the AdS/CFT correspondence can be used to
study form factors and their influence on $WW$ scattering.
One way to compute the form factor is based on the overlap integral
of onshell and offshell profiles of the states involved in the interaction.
This technique has been successfully used in AdS/QCD
models~\cite{Brodsky:2007hb,Hong:2004sa}, although \cite{Hong:2004sa}
suggests that this approach may not give trustworthy results at high energy
since expected results based on conformal scaling cannot be reproduced.
Furthermore, unlike previous applications, large brane kinetic terms in our
model can cause considerable deviation at low energy where the profile
distribution within the IR brane thickness is important.

Nonetheless let us consider the form factors obtained from the profile overlap
integral with the details of the derivation given in Appendix B. By performing this
simpler computation a preliminary phenomenological analysis of possible deviations
from the Standard Model can be studied. At high energy we find that the $E^4$
divergence in amplitudes of the $s,t$-channel $Z,\gamma$ exchange graphs are
sufficiently suppressed, as can be seen in the $WWZ$ form factors depicted in
Figs.~\ref{spacelikeWWZ} and \ref{timelikeWWZ} (see Appendix B). This is
because they involve a three-point vertex $WWZ$, $WW\gamma$ with an
offshell $Z,\gamma$ giving rise to a form factor which falls off as $\sim q^{-2}$ at
high energy, where $q$ is the transferred momentum carried by the intermediate
$Z,\gamma$. Concretely, we find that a good analytic approximation
for the $WWZ, WW\gamma$ form factor at low energy is
\beq
    F(q^2)\simeq\frac{m_Z^2}{m_Z^2+q^2}.
    \label{simpleff}
\eeq
This behavior agrees well with `vector-meson pole dominance' which follows
as a general result of a confining gauge theory with a gravity dual (although
large brane kinetic terms, as in our case, are not assumed)
and is also compatible with QCD data~\cite{Brodsky:2007hb,Hong:2004sa}.

However as we will see in the next section, LEP and Tevatron bounds on
the trilinear gauge boson vertex requires the form factor to be constant
for a larger energy range than the naive prediction (\ref{simpleff}) based on
the overlap integral method. As mentioned above this caveat exists
because the overlap integral method does not include the effects of the brane
thickness which should give a large correction to this prediction at low $q$ and
give a larger constant region well above $m_Z$. While an exact calculation
needs to be done we expect this behavior to follow from the fact that although
in the 4D description the $W,Z$-bosons are special resonances much lighter
than the IR scale, they are still composites of constituents confined at $\sim z_{IR}^{-1}$.
So a large deviation from point-like behavior should not occur well below $z_{IR}^{-1}$.
As discussed earlier from the 5D gravity perspective the $W,Z$-bosons are
similar to those in the original RS1 model--they are mostly localized on the IR brane
with a small profile that leaks into the bulk. Therefore well below
$z_{IR}^{-1}$ the bulk effect should be negligible and the gauge theory
on the IR brane should be a good effective description.

Furthermore using the overlap integral method the contact interaction does not
obtain a $q$-dependent form factor suppression because there is no offshell
transferred momentum involved. However a $q$-dependent form factor is in general expected
from the effective Lagrangian for composite vector boson interactions. This suggests that a more
comprehensive method to compute the three-point and four-point form factors is
to use 5D propagators to compute the three-point and four-point correlation
functions. This is, of course, a more involved calculation although related results
exist in the literature~\cite{Agashe:2005dk,Konyushikhin:2009kq}. A detailed
calculation will be postponed for a separate study.

\subsection{Unitarity of WW scattering}

As we noted in the last subsection the sum over both the contact and $s,t$-channel graphs
in our simplified computation (using the overlap integral method) is approximately
equivalent to the contact graph contribution alone with leading divergence
$E^4$ , since the $s,t$-channel graphs are suppressed. In fact this failure to cancel the
$E^4$ term is expected because gauge invariance is not exact in our composite model.
Compositeness induces energy-dependent form factors in the vertices and therefore
introduces terms forbidden by a fundamental gauge symmetry, ruining the usual gauge
cancelation mechanism. The cancelation is maximally violated using the overlap integral
method since it is not sensitive to energy-dependence in the contact interaction. It
therefore can be used to obtain a lower bound on the scale where tree-level unitarity breaks
down. Consider the leading divergence term for the contact graph amplitude:
\beq
    i\mathcal{A}^{(4)}(s,\cos\theta)=\frac{ig^2}{8 m_W^4}(3-\cos^2\theta)s^2,
\eeq
which leads to the $J=0$ partial wave amplitude
\beq
    a_0=\frac{1}{32\pi}\int^1_{-1} d(\cos\theta)\,\mathcal{A}^{(4)}(s,\cos\theta) =
    \frac{g^2}{48\pi m_W^4}s^2.
\eeq
Tree-level unitarity gives a bound on $a_0$ via the optical theorem,
namely: $|{\rm Re}~a_0|\leq\frac{1}{2}~$\cite{Gunion:1989we}. Using this,
we estimate the scale of tree-level unitarity breakdown to be $\sim300$\,GeV.
Again this bound assumes that the $E^4$ term from the contact interaction
is not suppressed compared to the $s,t$ channel graphs near 300 GeV
which are negligible--a drawback of the overlap integral method. It is expected that
a more accurate method which is sensitive to the energy dependence in the contact
interaction can delay the unitarity breakdown scale to be near the ${\cal O}(\rm TeV)$
confinement scale. A form factor suppression may cause a significant deviation from the
SM model prediction, where a faster growth of the $WW\rightarrow WW$
amplitude at low energy can lead to distinctive signals at the LHC, as we
will consider in the next section.

A natural question that remains to be answered is what happens at
high energy to eventually help restore unitarity? In analogy to hadron scattering 
in QCD, we expect two types of processes as the energy grows: hard 
elastic scattering and deep inelastic scattering (DIS).
High energy scattering of composite states depends on the physics of the 
underlying constituents. Even though at present we are unable to exactly specify 
the dual 4D gauge theory, some behavior of the underlying constituents 
(or `partons') of the $W,Z$-boson composite states (or `hadrons') and their influence 
on composite scattering at high energy can be ascertained based
on the gauge/string duality. 

As shown in~\cite{Polchinski:2001tt}, elastic scattering amplitudes for vector 
hadrons at large `t Hooft coupling fall as $E^{-2}$. This form factor suppression
can be intuitively understood by noting that at large momentum transfer $q$,
the entire hadron must shrink to a smaller size $\sim q^{-1}$ to scatter 
elastically, leading to a power-law suppression determined by the scaling of the 
wavefunction. This suggests that $W$-bosons in our emergent model undergo 
a similar process in the elastic scattering region where they shrink to a size
$q^{-1}$  and obtain a similar suppression. From the effective field theory
point of view, below the IR cutoff scale where only massless string states are relevant,
such a form factor should be calculable based on the 5D gravity model via 
three-point and four-point correlation functions of 5D massless gauge fields 
(massless string states). Again we postpone a detailed study to future work.

In the DIS region the $WW$ scattering amplitude becomes sensitive to 
short-distance physics associated with the underlying constituents. 
Analogous to QCD, two factors are relevant in this region to determine the 
scattering amplitude: the constituent-level scattering amplitude, and the 
structure function characterizing the distribution of hadron constituents.
The UV behavior of scattering constituent partons is expected to be soft 
both from the analogy with quark/gluon scattering in an asymptotically-free 
theory like QCD and the behavior of gluon scattering amplitudes in 
strongly-coupled CFTs~\cite{Alday:2007hr, McGreevy:2007kt}.
However there is a substantial difference from QCD as shown in 
Ref.~\cite{Polchinski:2002jw}: due to the large `t Hooft coupling, parton splitting 
is quite substantial for partons carrying a moderate Bjorken $x$. This means 
that there are no partons inside hadrons and causes the scattering to be 
dominated by color-neutral objects, which are the hadrons themselves. 
Partonic scattering eventually occurs below exponentially small $x\sim e^{-\sqrt{gN}}$ 
where the structure function becomes $q^2$ independent. Hence in the 
moderate $x$ region the whole hadron can only scatter `coherently'. One way 
this can occur is when the parent hadron splits into two pieces with each sub-hadron 
shrinking to a size of order $q^{-1}$, which eventually scatter and then rejoin to 
form the parent hadron. Therefore for moderate $x$ the effective `constituent' 
or `scattering unit' is sub-hadron whose structure function is suppressed at high 
$q$ due to the shrinking effect, similar to the form factor in the elastic region.

Although a careful study is needed to precisely ascertain how $WW$ 
unitarization occurs in our emergent model, it is promising that it already 
contains features such as form factor suppression ($\sim q^{-2}$), and 
UV soft parton scattering. Such a picture is not too dissimilar from that 
encountered in QCD. For example, an interesting analogy is to again consider 
$\rho$-meson scattering at high energy. Without knowing that they are 
composites of quarks and gluons, we might worry about tree-level unitarity 
when treating them as massive gauge bosons. But as is well known near
$\Lambda_{QCD}$, unitarity is eventually restored by partonic level
physics. Similarly, even though our model differs from QCD, it is
dual to a strongly-coupled gauge theory at large 't Hooft coupling
where similar effects could occur. In fact the non-QCD feature of
$WW$ deep inelastic scattering would be interesting to further
explore from the 5D string theory.

Finally we summarize the expected high-energy DIS behavior of $WW$
scattering in our model using the results in \cite{Polchinski:2002jw}. The 
total cross section $\sigma_T$ is given by
\beq
     \sigma_T = \sigma_H (e^{-\sqrt{g N}}\lesssim x \lesssim 1) +
     \sigma_P(0\lesssim x \lesssim e^{-\sqrt{g N}}),
\eeq
where $\sigma_H (\sigma_P)$ involves hadronic (partonic)
scattering. Assuming coherent scattering we can write
\beq
     \sigma_H=\int_{e^{-\sqrt{g N}}}^1 dx_1 dx_2 \, F(x_1,q^2) F(x_2,q^2)
     \sigma (WW\rightarrow WW),
\eeq
where $\sigma(WW\rightarrow WW)$ is the elastic scattering
cross section and $F(x,q^2)$ is the sub-hadron distribution
function. In particular for a vector boson, $F(x, q^2) \propto
q^{-2}$ according to conformal scaling~\cite{Polchinski:2002jw}
causing the hadronic cross section to fall sufficiently fast as the
energy grows. The parton cross section
\beq
       \sigma_P=\sum_{a,b}\int_0^{e^{-\sqrt{g N}}} dx_1dx_2\, f_a(x_1,q^2)f_b(x_2,q^2)
       \sigma(ab\rightarrow F) {\cal B}(F\rightarrow WW),
\eeq
where $f(x,q^2)$ is the parton distribution function and is
essentially $q^2$ independent~\cite{Polchinski:2002jw}. In this
region the branching ratio ${\cal B}(F\rightarrow WW)$ should give
sufficient suppression for $WW$ outgoing states. Therefore the large
`t Hooft coupling causes the scattering in the high-energy
region to be dominated by hadronic scattering ($\sigma_H$), while at
extremely small $x$ (nearly collinear scattering) we have partonic
scattering in the inelastic region.

\section{Collider Constraints and Signatures}

\subsection{Anomalous couplings}

The most important, generic phenomenological consideration of emergent electroweak symmetry breaking is the momentum dependent form factors that are induced in multi gauge boson interaction vertices.  It has been recognized for some time that composite gauge bosons can give rise to anomalous couplings amongst themselves~\cite{AGC}, leading to testable phenomena~\cite{Early Composite Pheno}.

Our primary task is to establish the viability of the theory when confronting the data that already exists. Since form factors start deviating with respect to the SM at higher energies, it is most expedient to compare the well-measured observables involving gauge bosons in the high-energy frontier to our theory.  These observables include $e^+e^-\to W^+W^-$ at the LEP2 collider, and $p\bar p\to W^\pm W^\mp,W^\pm Z$ at the Tevatron.
The first two of these processes involve the three-point interactions $\gamma W^+W^-$, and all three involve  the three-point interaction $ZW^+W^-$. Therefore, these observables are sensitive to  deviations in those three-point couplings.

To proceed, we must establish a notational framework that allows easy comparison to the reported experimental results. Deviations in triple gauge boson vertices
 from their SM values are often presented in the formalism of~\cite{Hagiwara:1986vm}:
\beq
\frac{{\cal L}_V}{g_{WWV}}=ig_1^Z(W^\dagger_{\mu\nu}W^\mu V^\nu-W^\dagger_\mu V_\nu W^{\mu\nu})+ik_VW_\mu^\dagger W_\nu V^{\mu\nu},
\eeq
where $g_1^Z=k_V=1$ at tree-level, $g_{WW\gamma}=-e$ and $g_{WWZ}=-e\cot\theta_w$.
It is convenient to define deviations from the SM, or `anomalous couplings', $\Delta g_1^Z$ and $\Delta k_V$ where $g_1^Z\equiv 1+\Delta g_1^Z$ and $k_V\equiv 1+\Delta k_V$.

As we see from the current limits summarized in Table~\ref{tgv limits}, deviations of only a few percent are tolerated by LEP2. Regarding Tevatron limits, a SM coupling is altered by a form factor suppression function $F(q^2)$ where $q^2$ is the momentum squared flowing into the vertex. Limits are set on $F(q^2)$ by replacing triple gauge boson vertex interaction couplings $g_{SM}$ with $g_{SM}F(q^2)$ and then computing the expected cross-section. Of course, at a lepton collider, the $e^+e^-\to Z^*\to W^+W^-$ cross-section for example probes the form factor simply at the invariant center of mass energy of the collider $F(q^2)\to F(s)$, where $s=(209\gev)^2$ at LEP2.

At the Tevatron and LHC,  the expected cross-section is an integral over the differential cross-section at many different invariant masses $q^2$. For example, for $p\bar p\to WZ$ at the Tevatron,
\bea
\frac{d\sigma}{dq^2} & = & \frac{1}{s}\sum_{a,b} \sigma_{ab}(q^2)
 \int_{q^2/s}^1\frac{dy}{y}\left[ f_{a/p}(q^2/sy)f_{b/\bar p}(y)+(a\leftrightarrow b)\right],
 \label{tevatron rho}
 \eea
 where $a,b$ are partons of the hadrons, $\sigma_{ab}(q^2)$ is the cross-section of $ab\to WZ$, and $s$ is the center of mass energy of the $p\bar p$  collisions. The $W^*WZ$ coupling within the computation for $\sigma_{ab}(q^2)$ is itself $q^2$ dependent due to the form factor suppression (and also renormalization group improvement, but that is subdominant here), and thus $F(q^2)$ gets sampled over many different values.
 If the expected  integrated cross-section is  outside the 95\% CL interval quoted by experiment, the form factor is said to be ruled out.

Of course, it is not convenient to speak abstractly of ruling out functions $F(q^2)$. Rather, it is more convenient to narrow $F(q^2)$ to a motivated subclass of functions with few parameters, and then constrain the parameters. With this in mind, and with guidance from previous theory papers (see, e.g.,  Eq.~(11) of~\cite{Hagiwara:1989mx}), the collaborations generally define $F(q^2)$  in terms of two parameters, $\Delta g$ and $\Lambda$, where the form factor plays the role of changing some suitably normalized SM coupling (such as $g_1^Z=1$) into
 \beq
 g_{SM}\Longrightarrow g_{SM}+\frac{\Delta g}{(1+q^2/\Lambda^2)^2},
 \label{eq:gdipole}
 \eeq
 where $g_{SM}=1$ or $0$ or whatever value is appropriate.    This is the so-called ``dipole form factor", and it is the convention by which experimental groups search for deviations. CDF and D0 often set $\Lambda=2\tev$ and then quote 95\% CL intervals on the $\Delta g$ anomalous couplings, as is presented in Table~\ref{tgv limits}.  At the LHC, it is customary to choose $\Lambda=10\tev$ when quoting expected sensitivity intervals on the anomalous couplings.

\begin{table}[t]
\begin{center}
\begin{tabular}{c|cccc}
\hline
                           & LEP2 & D0 ($1.1\xfbinv$) & CDF ($1.9\xfbinv$) & LHC ($30\xfbinv$) \\
 \hline\hline
 $\Delta g_1^Z$ & $(-0.051,0.034)$ & $(-0.14,0.34)$ & $(-0.13,0.23)$ & 0.0053 \\
 $\Delta k_\gamma$ & $(-0.105,0.069)$ & $(-0.51,0.51)$ & --- & 0.028\\
 $\Delta k_Z=\Delta g_1^Z$ & --- & $(-0.12,0.29)$ & $(-0.76,1.18)$ & 0.058\\
 \hline
 \end{tabular}
 \end{center}
 \caption{Limits on deviations of $g_1^Z$ and $k_V$ from their SM values. LEP2 results are taken from~\cite{Alcaraz:2006mx}, CDF and D0 results from~\cite{Iashvili:2008dj}. The CDF,  D0 and LHC results assume that there is a dipole form factor suppression for each anomalous coupling that scales as $\Delta g/(1+q^2/\Lambda^2)^2$. The limits on each $\Delta g$ in the table for CDF and D0 are derived by assuming $\Lambda=2\tev$. The LHC column is taken from the ``ideal case" limit of Atlas in~\cite{ATLAS TDR}, and assumes $\Lambda=10\tev$. All limits are at the 95\% CL.}
 \label{tgv limits}
 \end{table}

\subsection{Form factor viability of emergent electroweak symmetry breaking}

A good approximation to the form factor of emergent theory is
\beq
F(q^2)=\left\{
\begin{array}{l} 1,~~{\rm for}~q^2<\Lambda_C^2 \\
(1+\Lambda_C^2/\Lambda_{EW}^2)/(1+q^2/\Lambda_{EW}^2),~~{\rm for}~q^2>\Lambda_C^2\end{array}\right.
\label{eq:composite ff}
\eeq
where $\Lambda_{EW}$ and $\Lambda_C$  are taken to be free parameters. By the overlap integral
method $\Lambda_{EW}=m_Z$ is a low scale in comparison to typical interaction energies of current high-energy colliders. Therefore, this low scale of $\Lambda_{EW}$ induces significant suppressions at accessible collision energies. The more precise LEP2 limits rule out the model unless $\Lambda_C$ is greater than LEP2 center of mass energy, i.e., $\Lambda_C>209\gev$.

\begin{figure}[ttt]
\begin{center}
\includegraphics[angle=0,width=0.8\textwidth]{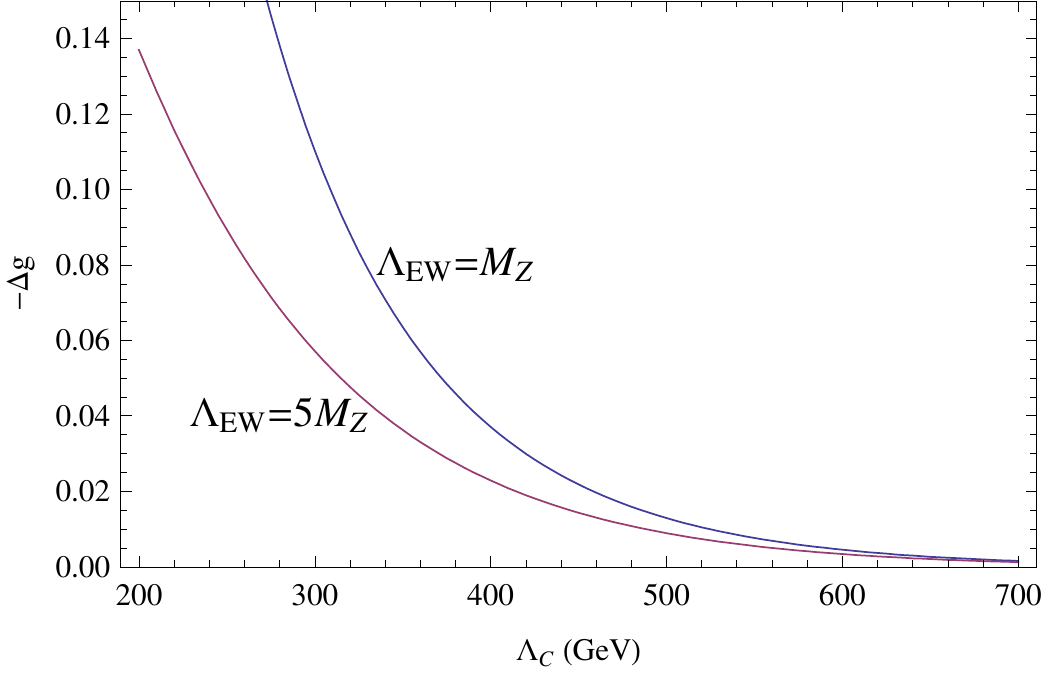}
\caption{ Contours of equal cross-section for $p\bar p\to W^+W^-$ in the $-\Delta g$ vs.\ $\Lambda_C$ plane for two different values of $\Lambda_{EW}$.  This plot enables a direct comparison between limits experimentalists obtain on $\Delta g$ after applying their form factor of Eq.~\ref{eq:gdipole} with $\Lambda=2\tev$, and the parameters of our form factor Eq.~\ref{eq:composite ff}.  The first line with $\Lambda_{EW}=m_Z$ is the value from the overlap integral method.  Tevatron limits require that $-\Delta g<0.13$ which implies that $\Lambda_C>285\gev$ at 95\% CL when $\Lambda_{EW}=m_Z$. \label{fig:tevatron dg}}
\end{center}
\end{figure}

We now wish to estimate what constraints Tevatron data puts on $\Lambda_C$. There is a challenge in doing this, since the form factor of (\ref{eq:composite ff}) is substantially different than the form factor
of (\ref{eq:gdipole}) that the experimentalists use to quote limits on the anomalous couplings.
The method we use to compare is simply to obtain the total cross-section for some given $\Delta g$ in the experimentalists form factor definition (with $\Lambda=2\tev$) and then find what values of $\{\Lambda_{EW},\Lambda_C\}$ match that cross-section and draw a cross-section equivalence plot. In
Figure~\ref{fig:tevatron dg} we show this correspondence of $\Delta g$ with $\Lambda_C$ for two different values of $\Lambda_{EW}$: $\Lambda_{EW}=m_Z$, which is the value obtained using the overlap integral method, and the somewhat higher value of  $\Lambda_{EW}=5m_Z$.  Of course, higher $\Lambda_{EW}$ and higher $\Lambda_C$ correspond to smaller  magnitudes of anomalous couplings $\Delta g$.

We see from the plot that the 95\% CL limit of $-0.13$ (see Table~\ref{tgv limits}) for the anomalous coupling of triple gauge boson vertices corresponds to $\Lambda_C=285\gev$ ($205\gev$) if $\Lambda_{EW}=m_Z$ ($5m_Z$). These values are thus the lower limits on $\Lambda_C$ from Tevatron analyses.
A reasonable estimate for $\Lambda_C$ is near the TeV scale below which all couplings are SM-like.
Violations of perturbative unitarity only happen at $q^2>\Lambda_C^2$ where new degrees of freedom associated with the compactification scale come in to unitarize the amplitude. In this case, strongly coupled $VV\to VV$ scattering becomes an interesting tool of discovery for these theories.

\subsection{Large Hadron Collider}

The expected sensitivities of anomalous triple gauge boson vertices at the LHC is more than an order of magnitude better than Tevatron capabilities.  This can be seen from the expected sensitivities at the LHC in $30\xfb^{-1}$ presented in Table~\ref{tgv limits}.
If a signal for beyond the SM physics does develop in vector boson scattering at the LHC, there are many options to study the detailed underlying theory. Measuring all possible observables associated with vector boson final states will be part of the physics programme at the LHC in any event, and one would be able to study quantitatively all the changes that occur. These studies can be broken up into two categories, vector boson fusion processes and diboson production processes.

$VV\to VV$ scattering can be separated from other modes of generating $VV$ final states by looking for a few characteristics of the final state associated with these modes. The most important feature is that the initial state  vector bosons must arise by radiating off incoming quarks of the proton, and thus there will be two extra jets that are at high  rapidity accompanying the event. This is why $VV\to VV$ is often expressed as the equivalent $pp\to VVjj$, where these last two `tagging' jets are measured. To further isolate this signal over other potential backgrounds, such as $t\bar t$ production, it is required that there is very little jet activity in the central region. This is characteristic of the signal since no color exchanges across the wide central rapidity occur, and emission of QCD radiation is suppressed.  These characteristics have been understood for some time now~\cite{VVjj Refs}. An example quantification is given by the ATLAS collaboration, who has chosen for some samples that $\Delta \eta=5$ for the two tagging jets (far separated) with energies greater than $300\gev$,  and that no other more central jet exists with $p_T>30\gev$~\cite{AtlasBook}.

It is this $VV\to VV$ vector boson scattering that causes concern for unitarity discussed earlier. If an effective form factor suppression for quartic gauge boson vertices scales as $1/q^4$ and the would-be partial wave amplitude  scales as $q^4$ (see discussion in Section~\ref{wwsection}), the resulting composite amplitude scales as $|A|^2\sim q^4/q^4=q^0$. Therefore, the
 total cross-section for $VV\to VV$ scales with $|A|^2/q^2\sim 1/q^2$ as the SM rate does, and the resulting differential cross-section could be similar in value to the SM. It is unlikely that it will be the same, and thus studies of this mode are extremely useful to see the precise differences. The differences are not computable at this time, but measurement will have its own enduring value as theory catches up.

If the form-factor scale $\Lambda_C$ is well above the TeV scale, then it becomes important to consider the small deviations of the coupling from the assumed value of 1 (i.e., its SM value) in our form factor for $q^2<\Lambda_C^2$.  Small deviations are to be expected, but there is no obvious functional form to choose to study this. Therefore, any reasonable functional form that is descriptive to changes of observables from SM values will do. A convenient choice would simply be the choice made by the experimental collaborations, and the results in Table~\ref{tgv limits} are to be consulted for LHC sensitivity of these deviations.

The second kind of process is diboson production, which we define here to mean all underlying two to two processes initiated by quarks that generate vector boson pairs. It is in these studies that the triple gauge boson vertices can be measured and compared to SM values. As we mentioned above, our  model may even have very large deviations compared to LHC sensitivities. Nevertheless, we wish to make some further comments about how to study the deviations if they occur, and what qualitative features would develop in the observables if this model is a correct description of nature.

A fruitful approach to organizing observables is to separate the processes that mostly involve multiple gauge boson interactions versus those processes that do not. For example, the production of $W^\pm Z$ at the LHC is primarily through $s$-channel $W^*$, and thus its rate is highly dependent on the details of the $WWZ$ vertex. On the other hand, $ZZ$ production at the LHC is primarily initiated by $t$-channel quark exchange diagrams and the production cross-section is mostly dependent on the $Z\bar q q$ interaction vertex (assuming
this vertex has negligible form factor suppression)~\cite{Brig:2007}. At high invariant mass energies, greater than about $\Lambda_C$, the $ZZ$ production cross-section in our model should remain similar to that of the SM, whereas the $W^\pm Z$ production cross-section should diminish rapidly.  Thus, a useful signature would be the ratio $\sigma(WZ)/\sigma(ZZ)$ as a function of center of mass energy of the final state vector bosons. We expect that well above $\Lambda_C$ this ratio will be significantly diminished compared to the SM prediction of
\beq
\frac{\sigma(WZ\to 3l+\nu)}{\sigma(ZZ\to 2l+2\nu)} = 1.7\pm 05,
\eeq
in the induced three to two lepton ratio. This ratio is obtained from~\cite{AtlasBook} after cuts applied.

As we have seen, few of the observables in the LHC regime can be computed precisely at this time due to the complexity in determining couplings at high energies in this theory. Despite this, we know that several generic features must come about, and these features can be verified by experimental measurements: vector boson production at high invariant mass will be altered by non-perturbative dynamics one way or another, through dramatic suppression factors or through new dynamics unitarizing amplitudes, and observables that are supported mostly by triple gauge boson vertices will show a differential suppression in rate compared to other observables $q^2>\Lambda_C^2$.

Finally, we wish to remark that the KK photon described earlier is likely to be the lightest exotic state in the spectrum. Its phenomenological signatures are very similar to $Z'$ physics well studied in the literature. The discovery reach depends of course  on the precise couplings, which are not determined at this time but are expected to be ${\cal O}(1)$ in electroweak strength.  From a variety of $Z'$ theories with electroweak coupling strength one can estimate that direct limits from Tevatron should be $M_{KK}\gsim 800\gev$ and direct limits from LHC should be $M_{KK}\gsim 3\tev$ after $10\xfb^{-1}$ of
data~(see e.g.\ Figs.~1.6 and 1.7 of \cite{Rizzo:2006nw}).  If the coupling drops well below electroweak strength, which may occur from some exotic choices of fermion profiles and gauge kinetic terms on the brane,   decoupling from collider observables happens rapidly and a surprisingly low mass scale -- even tens of GeV -- could be allowed phenomenologically (cf.\ \cite{Kumar:2006gm}), although this extreme is unlikely from our theory point of view.

\section{Conclusion}

We have presented a model of electroweak symmetry breaking in a warped dimension
where electroweak symmetry is broken at the Planck scale. The masses of the $W,Z$ bosons
result from the breaking of conformal symmetry and do not rely on a Higgs mechanism.
Large brane kinetic terms are responsible for generating an anomalously light first KK mode
that can be identified with an electroweak gauge boson, while simultaneously allowing the
higher KK modes to be at the TeV scale.

Interestingly, by the AdS/CFT correspondence this model is dual to a strongly-coupled CFT where
the $W,Z$ bosons are identified as composite states. In this way there is no fundamental electroweak
symmetry in our model and electroweak symmetry breaking emerges in the IR. This realizes an
old idea of mimicking the electroweak gauge bosons with the $\rho$-mesons in QCD, except that in our
model the 4D theory is always strongly-coupled. Furthermore via the gravity dual we are able to quantitatively check consistency with  electroweak precision tests. In particular we find reasonable fits to the $S$ and $T$ parameters as well as show that the $V$ parameter is likely to be small. A novel feature of our setup is that there is a custodial symmetry (a global SU(2) symmetry in the dual CFT) which protects the $T$ parameter and is
akin to isospin symmetry in QCD.

The composite nature of the $W,Z$ bosons gives rise to energy-dependent form factors and suggests distinctive signatures at the LHC. Partly motivated by the form factor derived by the profile overlap integral method we have
presented a more general form factor that characterizes emergent electroweak symmetry breaking. This form factor has been used to analyse constraints arising from LEP2 and the Tevatron, updating previous analyses and shows
that there is still enough parameter space to allow for composite gauge bosons. In fact the LHC has potential to discover deviations in triple gauge boson vertices and we have suggested ways to organize observables in order to optimize future searches.

Our emergent model is by no means complete. Although we have identified how fermion masses
can be incorporated in a straightforward manner we have not performed a detailed analysis and this could
affect the model in substantial ways. Similarly a better understanding of unitarity in $W$-boson scattering is
needed. This most likely requires computing the three- and four-point correlation functions in 5D to
confirm the form factor suppression in the high-energy elastic scattering argued to help restore unitarity. In addition we relied on the brane thickness of the IR brane to argue for a smooth interpolating limit between a brane-localized field and a bulk field. This is a crucial aspect of our model and a thorough investigation of how exactly this affects the fermion and form factor calculations remains to be done. In addition there are interesting questions such as the nature of the underlying gauge theory and string theory realization as well as an intriguing Seiberg duality. 
Even though there are remaining issues and questions, our emergent model does provide a glimpse of how 
composite $W,Z$ bosons can be made compatible with experiment in a framework that can be used to further 
develop this idea. It could be that electroweak symmetry breaking arises from a deeper level of substructure 
underlying the SM, where there is no Higgs sector and electroweak gauge symmetry is not even fundamental.
The LHC will soon let us know.

\section*{Acknowledgements}
We thank Brian Batell, David Morrissey, Alex Pomarol, Marco Serone, and
Raman Sundrum for helpful discussions. Y.C. is supported by the Harvard
Center for Fundamental Laws of Nature. T.G. is supported by the Australian
Research Council and acknowledges the Jefferson Lab at Harvard
University for hospitality where part of this work was done.
J.W. is supported in part by the United States Department of Energy.
The authors also acknowledge the Kavli Institute for Theoretical Physics at
Santa Barbara where some of this work was undertaken.

\appendix
\def\theequation{\thesection.\arabic{equation}}
\setcounter{equation}{0}
\section{Alternative derivation of a Light Kaluza-Klein mode with Brane Kinetic Terms}

In this Appendix we present a more transparent way to see
how a light collective mode appears when a brane kinetic term is added.
As will be shown a brane kinetic term induces significant renormalizing
and mixings of the original Kaluza-Klein kinetic term (i.e. before adding brane kinetic terms),
which after canonically normalizing the kinetic term and diagonalizing the mass
matrix gives rise to a suppression factor in the mass of one special mode.

For simplicity we will consider the case where the boundary conditions are $(-+)$.
Then after adding an IR brane kinetic term we
can study how the kinetic terms and mass matrix of the Kaluza-Klein modes changes.
This is exactly the same situation for the $W$-boson in our model, and similarly
expect the qualitative features to generalize to the more complicated case of the
$Z$-boson with mixed boundary conditions.

The 4D Lagrangian of the KK modes with $(-+)$ boundary conditions is given by:
\beq
    \mathcal{L}_4= \sum_{n=1}^\infty-\frac{1}{4}(F_{\mu\nu}^{(n)}(x))^2-\frac{1}{2}m_n^2(A_\mu^{(n)}(x))^2,
\eeq
where the Kaluza-Klein masses $m_n\neq0$, since with $(-+)$ boundary conditions there is no
massless zero mode. Note that in the KK basis $A_\mu^{(n)}$, the kinetic terms of all KK modes are
canonically normalized and the Kaluza-Klein mass matrix is diagonal.
Denoting the 5D profile for the $n$th KK mode by $f_n(z)$, it is known that the IR overlap
$f_n(z_{IR})\sim\sqrt{2k}$ is approximately universal for all KK modes \cite{Pomarol:1999ad}.

Let us now introduce an IR brane kinetic term of the form:
\beq
    \triangle\mathcal{L}_4=-\sum_{n,k=1}^\infty \frac{1}{4}\zeta_{IR} F_{\mu\nu}^{(n)}(x) F^{\mu\nu (k)}(x) f_n(z_{IR})f_k(z_{IR}),
    \label{braneKE}
\eeq
where $\zeta_{IR}$ is a constant in units of $k^{-1}$. The introduction of (\ref{braneKE}) contains additional
mixings in the kinetic terms of the original KK modes in the 4D Lagrangian. Thus in the KK basis $A_\mu^{(n)}$,
instead of an identity matrix, the kinetic energy matrix becomes
\beq
               \left(\begin{array}{ccc}
                      1+2\zeta_{IR}k & 2\zeta_{IR}k &\ldots \\
                      2\zeta_{IR}k & 1+2\zeta_{IR}k & \ldots \\
                       \vdots& \vdots & \ddots \\
                    \end{array}
                  \right).
\label{kematrix}
 \eeq
While in principle this is an infinite-dimensional matrix, it is usually truncated as a
finite $N\times N$ matrix where $N\simeq\Lambda_{IR} z_{IR}\sim 10-100$
characterizes the number of KK modes below the local cutoff scale, $\Lambda_{IR}$.
It is easy to check that a truncated version of the matrix (\ref{kematrix}) can be diagonalized with eigenvalues:
\beq
\left(
  \begin{array}{cc}
    {\cal I}_{N-1,N-1}& 0 \\
    0 & 1+Na \\
  \end{array}
\right),
\eeq
where $a=2\zeta_{IR} k$ and ${\cal I}_{N-1,N-1}$ is $(N-1)\times(N-1)$ unit matrix.
Clearly all eigenvalues are one except for a special eigenmode which has a
large eigenvalue $1+Na$. In terms of the original KK basis this
special eigenmode is given by
\beq
A_\mu^{\prime(N)}=\frac{1}{\sqrt{N}}\displaystyle\sum_{n=1}^{N}A_\mu^{(n)}.
\eeq
This relation implies that this mode is like a `collective' mode containing an
equal contribution from each of the original KK modes $A_\mu^{(n)}$.
To canonically normalize the kinetic term of $A_\mu^{\prime(N)}$, we perform
a non-unitary rescaling and define the normalized mode
$A^{\prime\prime(N)}_\mu\equiv\sqrt{1+Na}\, A^{\prime(N)}_\mu$. As we will show it is
essentially this large rescaling factor associated with the collective
mode that leads to a large suppression factor for the mass term related
to the light mode after diagonalizing the KK mass matrix.

The next step is to diagonalize mass matrix. Analytical expressions can be obtained for the
case $N=3$. The transformation leading to the new basis $A_\mu^{\prime\prime(n)}$ with canonically normalized kinetic terms is
\beq
\left( \begin{array}{c}
    A_\mu^{\prime\prime(1)} \\
    A_\mu^{\prime\prime(2)} \\
    A_\mu^{\prime\prime(3)} \\
  \end{array}\right)=
\left(\begin{array}{ccc}
    -\frac{1}{\sqrt{2}} &0 &  \frac{1}{\sqrt{2}} \\
     -\frac{1}{\sqrt{6}} &  \sqrt{\frac{2}{3}} &  -\frac{1}{\sqrt{6}} \\
    \sqrt{\frac{1+3a}{3}} & \sqrt{\frac{1+3a}{3}} & \sqrt{\frac{1+3a}{3}} \\
  \end{array}\right)
\left(  \begin{array}{c}
    A_\mu^{(1)} \\
    A_\mu^{(2)} \\
    A_\mu^{(3)} \\
  \end{array}\right).
  \label{dptrans}
 \eeq
Using (\ref{dptrans}) the diagonal mass-squared matrix in the original basis can be rewritten in terms of the new basis $A_\mu^{\prime\prime(n)}$ to give:
\beq
\left(\begin{array}{ccc}
    \frac{1}{2}(m_1^2+m_3^2) &\frac{1}{2\sqrt{3}}(m_1^2-m_3^2) & \frac{-1}{\sqrt{6}}\frac{m_1^2-m_3^2}{\sqrt{1+3a}}  \\
    \frac{1}{2\sqrt{3}}(m_1^2-m_3^2)  &  \frac{1}{6} (m_1^2+4m_2^2+m_3^2) &
    \frac{-1}{3\sqrt{2}} \frac{m_1^2-2m_2^2+m_3^2}{\sqrt{1+3a}}   \\
  \frac{-1}{\sqrt{6}}\frac{m_1^2-m_3^2}{\sqrt{1+3a}}  & \frac{-1}{3\sqrt{2}} \frac{m_1^2-2m_2^2+m_3^2}{\sqrt{1+3a}}  &
  \frac{1}{3}\frac{m_1^2+m_2^2+m_3^2}{1+3a} \\
\end{array}\right).
\label{m2dp}
\eeq
The eigenvalues of the mass matrix (\ref{m2dp}) contain two eigenvalues of order $z_{IR}^{-1}$, as well as
a light mode with mass $\sim z_{IR}^{-1}/\sqrt{a}$. By increasing $N$ one can numerically check that these results
persist and there is always a light mode with mass:
\beq
    m_1^\prime\propto \frac{z_{IR}^{-1}}{\sqrt{\zeta_{IR} k}}~.
\eeq
Thus, up to an order one factor this result agrees well with the mass found in Section~\ref{sec5dmodel}
by directly solving the equations of motion with the boundary conditions. This shows that the existence of a light collective mode originates from a large wavefunction renormalization induced by kinetic mixing
from the boundary kinetic term.

\section{Form Factor Calculation via Profile Overlap Integral}

In this Appendix we present the details of computing form factors using the profile overlap integral.
As discussed in the main text we do not include the effects of large brane kinetic terms which
will likely modify the low $q$ behavior. Furthermore the overlap integral method does not give
momentum dependent form factors for the contact interaction. Therefore any deviation in the
$WW$ scattering amplitude using this method
is encoded in the form factors of $WWZ, WW\gamma$
vertices.  To be more precise, possible intermediate states in the
$s,t$ channel exchange graphs contain $\gamma,Z$-boson and higher KK
modes $\gamma^{(n)},Z^{(n)}$. These modes have a universal form
factor in their coupling to $WW$. In 5D models the form factor can be
calculated in terms of a wave function overlap integral.
An AdS/QCD example can be found in \cite{Brodsky:2007hb} where the pion electromagnetic form factor is
calculated by an overlap integral of an onshell pion profile $\Phi(z)$ and offshell photon profile $J(q^2,z)$,
where $q$ is the transferred momentum carried by the probe photon.

Therefore in the overlap integral external states or states whose compositeness is being probed--like the
incoming and outgoing $W$-bosons--are represented by an on-shell profile, while the probe as the intermediate
state like the exchanged $\gamma,Z$ is represented by a general offshell profile with $q^2$ dependence.
The offshell profile is inferred from the known onshell profile by simply replacing $m^2$ by $-q^2$ in the bulk solution.
Thus the general solution $f(q^2,z)$ for offshell modes following
from the bulk equation of motion (\ref{bulkeom}) is
\beq
   f(q^2,z)=N z(J_1(\sqrt{-q^2}z)+b Y_1(\sqrt{-q^2}z)),
   \label{bulksolp}
\eeq
where $N,b$ are arbitrary constants. Notice that depending on the sign of $-q^2$,
the solution can be divided into two qualitatively distinct regions: a timelike region
with $-q^2>0$, which includes the onshell case, and a spacelike region with $-q^2<0$.
In fact it is more convenient to define a real positive variable, $p\equiv\sqrt{-q^2}$
for the timelike region, while $p\equiv-i \sqrt{-q^2}$ for the spacelike region.

To determine the constants $N,b$ we require the general solution $(\ref{bulksolp})$
satisfy two conditions: it should match the onshell result at $-q^2=m^2$, and
be canonically normalized like an onshell mode\footnote{In the electroweak
precision analysis section
only the photon is strictly canonically normalized, while the normalization of $Z$
is close to $1$. But for the present purposes, it is sufficient to require canonical
normalization for all modes.}, so that all $p-$dependent changes
due to compositeness are represented by a vertex form factor. If canonical
normalization is not chosen then there is a form factor associated with the
modified propagator which is an equivalent but more complicated procedure.

These conditions can be satisfied by imposing the same IR boundary conditions
for the offshell mode with the proper substitution of $m^2\rightarrow-q^2$ and then
canonically normalizing the profile (equivalent to imposing a $p-$dependent value
as the Dirichlet boundary condition for the source field on the UV brane).
This leads to the form factor:
\beq
F_{WWZ}(q^2)=\frac{1}{N_Z(q^2)N_W^2}\left\{\left[ \int_{z_{UV}}^{z_{IR}}\frac{dz}{kz}
f^{L3}(q^2,z)(f_W(z))^2\right]+\zeta_Lf^{L3}(q^2,z_{IR})(f_W(z_{IR}))^2 \right\}.
\eeq
Notice that only the $f^{L3}$ component of the $Z$ profile is relevant since the interaction
comes from the kinetic term in the bulk Lagrangian $-\frac{1}{4}(F^{La}_{MN})^2$.
The onshell $W$-boson profile $f_W$ is approximately given by (\ref{wprofile}), and
$N_W$ is obtained by requiring the canonical normalization of $f_W(z)$. Using
(\ref{bulksol}) and (\ref{coeff}) $f^{L3}(q^2,z)$ is given by
\beq
     f^{L3}(q^2,z)=N_Z(q^2) z\left[J_1(\sqrt{-q^2}z)+b_n^{L3}(q^2)Y_1(\sqrt{-q^2}z)\right],
\eeq
where
\beq
      b_n^{L3}(q^2)=\frac{\zeta_Lk\sqrt{-q^2}z_{IR}J_1(\sqrt{-q^2}z_{IR})-
      J_0(\sqrt{-q^2}z_{IR})}{Y_0(\sqrt{-q^2}z_{IR})-\zeta_Lk\sqrt{-q^2}z_{IR}Y_1(\sqrt{-q^2}z_{IR})}.
\eeq
The normalization $N_Z(q^2)$ is given by
 \bea
N_Z^2(q^2)&=&\left[\int^{z_{IR}}_{z_{UV}}\frac{dz}{kz}(f^B(q^2,z))^2+(f^{L3}(q^2,z))^2\right]
+\zeta_L(f^{L3}(q^2,z_{IR}))^2+\zeta_Y(f^B(q^2,z_{IR}))^2\nonumber\\
&&\qquad\qquad\qquad+\frac{\zeta_Q}{g_{Y5}^2+g_{L5}^2}\left(g_{Y5}f^B(q^2, z_{UV})
+g_{L5}f^{L3}(q^2,z_{UV})\right)^2,
 \eea
where $f^B(q^2,z)$ still appears indirectly through the normalization condition. Like
$f^{L3}(q^2,z)$, an explicit expression for $f^B(q^2,z)$ is obtained from (\ref{bulksol}),
(\ref{coeff}) and (\ref{coeffalp}).

\begin{figure}
\begin{center}
        \includegraphics[width = 0.7\textwidth]{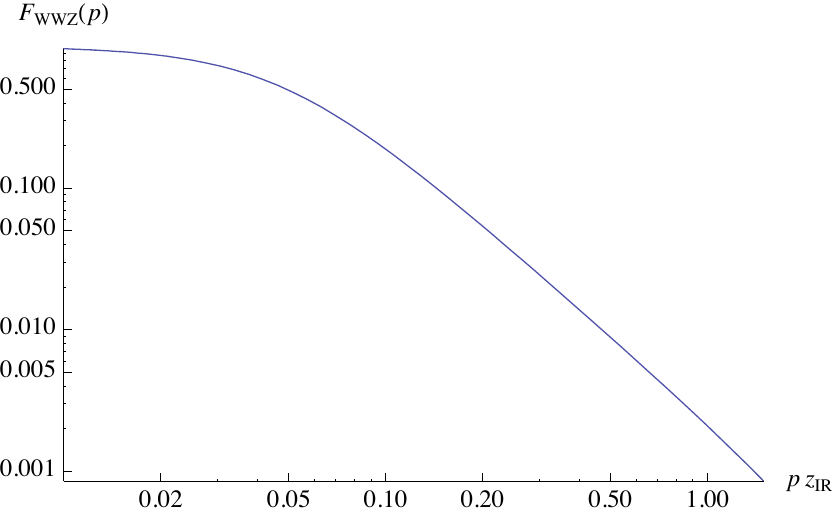}
\label{ptform}
\end{center}
\caption{The form factor $F_{WWZ}(p)$ in the spacelike region for $z_{IR}^{-1}$ = 1.8 TeV.}
\label{spacelikeWWZ}
\end{figure}

\begin{figure}
\begin{center}
        \includegraphics[width = 0.7\textwidth]{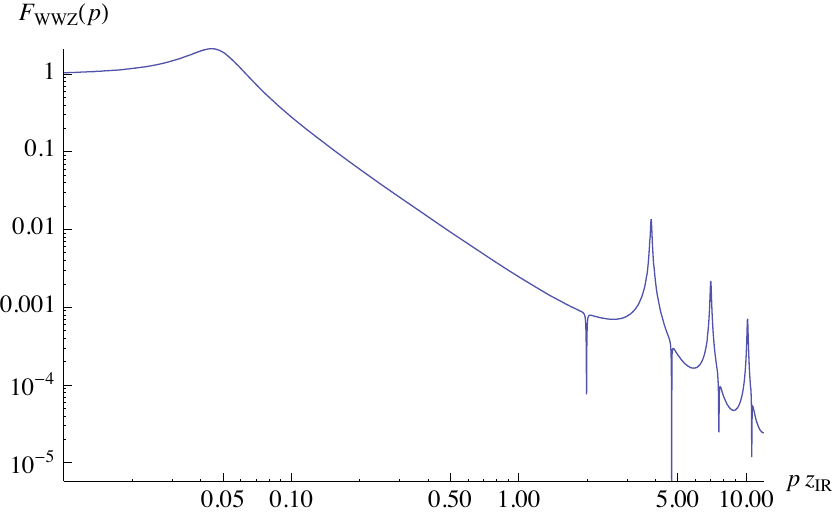}
\label{psform}
\end{center}
\caption{The form factor $F_{WWZ}(p)$ in the timelike region for $z_{IR}^{-1}$ = 1.8 TeV.
The pole structure corresponds to the photon and $Z$-boson KK modes.}
\label{timelikeWWZ}
\end{figure}

It is difficult to obtain an analytic expression for the form factor since the small argument
expansion for Bessel functions that was used for light onshell modes, is no longer valid
with a more general argument like $pz$.  Instead the form factor $F_{WWZ}(p)$ can be evaluated
numerically and then an analytic fit can be done in the low-energy region to obtain
an approximate analytic expression. The numerical results are shown in Figs.~\ref{spacelikeWWZ},
and \ref{timelikeWWZ} where the form factor is normalized so that $F(0)=1$.

In Figure~\ref{spacelikeWWZ} the form factor in the spacelike region monotonously
decreases as $p$ increases.  This behavior also occurs in the timelike region
in Figure~\ref{timelikeWWZ}, except that it is interspersed with periodic
peaks and troughs. The position of every peak and trough coincides exactly with the
KK mass poles: peaks are KK $Z$-bosons, while troughs are KK photons.
This form factor structure can be compared with that expected from a confining gauge
theory with a gravity dual, namely~\cite{Hong:2004sa}
\beq
     F_{ab}(q^2)=\displaystyle\sum_{n=1}\frac{f_ng_{nab}}{q^2+m_n^2},
\eeq
where $F_{ab}(q^2)$ is the form factor associated with a spin-1
current for two external hadrons $a,b$, with coupling $g_{nab}$
to an $n$th vector hadron state with mass $m_n$ and decay constant $f_n$.
This expression shows that in a confining gauge theory with large `t  Hooft
coupling, the form factor can be written as a sum over vector-meson poles.
Our results are consistent with this formal prediction:
a simple analytic fit in the low energy region ($Z$-pole dominant) for both the
timelike and spacelike form factor gives $m_Z^2/(m_Z^2 +p^2)$, which appears
as Eq.(\ref{simpleff}).

Note that although it is difficult to find a global fit for the structure of the time-like region,
this fit is good in the low-energy region (smooth out the peak at $Z$ pole) which
is monotonous and more relevant for LHC study. Interestingly a similar pole structure
was originally conjectured for QCD by Sakurai~\cite{Sakurai:1960ju,Hong:2004sa},
where it was suggested that the form factor of an isospin-hadron $H$ is given by the
$\rho$ meson pole: $F(q^2)\approx\frac{f_\rho g_{\rho HH}}{q^2+m_\rho^2}$.
For practical application, we can show that based on kinematics in high-energy
scattering the momentum transferred in the $t$-channel is mostly spacelike, while in
the $s$-channel it is always timelike.


\begin{thebibliography}{[99]}

\bibitem{Kaplan:1983fs}
  D.~B.~Kaplan and H.~Georgi,
  Phys.\ Lett.\  B {\bf 136}, 183 (1984).

\bibitem{Agashe:2004rs}
  K.~Agashe, R.~Contino and A.~Pomarol,
  Nucl.\ Phys.\  B {\bf 719}, 165 (2005)
  [arXiv:hep-ph/0412089].

\bibitem{Csaki:2003zu}
  C.~Csaki, C.~Grojean, L.~Pilo and J.~Terning,
  Phys.\ Rev.\ Lett.\  {\bf 92}, 101802 (2004)
  [arXiv:hep-ph/0308038].

\bibitem{AGC}
  L.~F.~Abbott and E.~Farhi,
  Phys.\ Lett.\  B {\bf 101}, 69 (1981).\\
  H.~Fritzsch, D.~Schildknecht and R.~Kogerler,
  Phys.\ Lett.\  B {\bf 114}, 157 (1982).\\
  T.~Kugo, S.~Uehara and T.~Yanagida,
  Phys.\ Lett.\  B {\bf 147}, 321 (1984).\\
  S.~Uehara and T.~Yanagida,
  Phys.\ Lett.\  B {\bf 165}, 94 (1985).\\
  M.~Bando, T.~Kugo and K.~Yamawaki,
  Prog.\ Theor.\ Phys.\  {\bf 73}, 1541 (1985).\\
  M.~Suzuki,
  Phys.\ Rev.\  D {\bf 37}, 210 (1988).\\
  M.~Bando, T.~Kugo and K.~Yamawaki,
  Phys.\ Rept.\  {\bf 164}, 217 (1988).

\bibitem{Bando:1984ej}
  M.~Bando, T.~Kugo, S.~Uehara, K.~Yamawaki and T.~Yanagida,
  Phys.\ Rev.\ Lett.\  {\bf 54}, 1215 (1985).

\bibitem{Gross:1997wg}
  D.~J.~Gross,
{\it  In *Liu, C.S. (ed.): Yau, S.T. (ed.): Chen Ning Yang* 147-162}.

\bibitem{Seiberg:1995ac}
  N.~Seiberg,
  Int.\ J.\ Mod.\ Phys.\  A {\bf 16}, 4365 (2001)
  [arXiv:hep-th/9506077].

\bibitem{Maldacena:1997re}
  J.~M.~Maldacena,
  Adv.\ Theor.\ Math.\ Phys.\  {\bf 2}, 231 (1998)
  [Int.\ J.\ Theor.\ Phys.\  {\bf 38}, 1113 (1999)]
  [arXiv:hep-th/9711200].

\bibitem{ArkaniHamed:2000ds}
  N.~Arkani-Hamed, M.~Porrati and L.~Randall,
  JHEP {\bf 0108}, 017 (2001)
  [arXiv:hep-th/0012148].

\bibitem{Randall:1999ee}
  L.~Randall and R.~Sundrum,
  Phys.\ Rev.\ Lett.\  {\bf 83}, 3370 (1999)
  [arXiv:hep-ph/9905221].

\bibitem{Amsler:2008zzb}
  C.~Amsler {\it et al.}  [Particle Data Group],
  Phys.\ Lett.\  B {\bf 667}, 1 (2008).

\bibitem{Georgi:2000ks}
  H.~Georgi, A.~K.~Grant and G.~Hailu,
  Phys.\ Lett.\  B {\bf 506}, 207 (2001)
  [arXiv:hep-ph/0012379].

\bibitem{Davoudiasl:2004pw}
  H.~Davoudiasl, J.~L.~Hewett, B.~Lillie and T.~G.~Rizzo,
  JHEP {\bf 0405}, 015 (2004)
  [arXiv:hep-ph/0403300].

\bibitem{Csaki:2005vy}
  C.~Csaki, J.~Hubisz and P.~Meade,
  arXiv:hep-ph/0510275.

\bibitem{Dvali:2000rx}
  G.~R.~Dvali, G.~Gabadadze and M.~A.~Shifman,
  Phys.\ Lett.\  B {\bf 497}, 271 (2001)
  [arXiv:hep-th/0010071].

\bibitem{Carena:2002me}
  M.~S.~Carena, T.~M.~P.~Tait and C.~E.~M.~Wagner,
  Acta Phys.\ Polon.\  B {\bf 33}, 2355 (2002)
  [arXiv:hep-ph/0207056].

\bibitem{Carena:2002dz}
  M.~S.~Carena, E.~Ponton, T.~M.~P.~Tait and C.~E.~M.~Wagner,
  Phys.\ Rev.\  D {\bf 67}, 096006 (2003)
  [arXiv:hep-ph/0212307].

\bibitem{Chacko:1999hg}
  Z.~Chacko, M.~A.~Luty and E.~Ponton,
  JHEP {\bf 0007}, 036 (2000)
  [arXiv:hep-ph/9909248].

\bibitem{Ponton:2001hq}
  E.~Ponton and E.~Poppitz,
  JHEP {\bf 0106}, 019 (2001)
  [arXiv:hep-ph/0105021].

\bibitem{Davoudiasl:2002ua}
  H.~Davoudiasl, J.~L.~Hewett and T.~G.~Rizzo,
  Phys.\ Rev.\  D {\bf 68}, 045002 (2003)
  [arXiv:hep-ph/0212279].

\bibitem{Sundrum:1998sj}
  R.~Sundrum,
  Phys.\ Rev.\  D {\bf 59}, 085009 (1999)
  [arXiv:hep-ph/9805471].

\bibitem{delAguila:2003bh}
  F.~del Aguila, M.~Perez-Victoria and J.~Santiago,
  JHEP {\bf 0302}, 051 (2003)
  [arXiv:hep-th/0302023].

\bibitem{Giddings:2001yu}
  S.~B.~Giddings, S.~Kachru and J.~Polchinski,
  Phys.\ Rev.\  D {\bf 66}, 106006 (2002)
  [arXiv:hep-th/0105097].

\bibitem{Grossman:1999ra}
  Y.~Grossman and M.~Neubert,
  Phys.\ Lett.\  B {\bf 474}, 361 (2000)
  [arXiv:hep-ph/9912408].

\bibitem{Gherghetta:2000qt}
  T.~Gherghetta and A.~Pomarol,
  Nucl.\ Phys.\  B {\bf 586}, 141 (2000)
  [arXiv:hep-ph/0003129].

\bibitem{adscftdict}
 N.~Arkani-Hamed, M.~Porrati and L.~Randall,
  JHEP {\bf 0108}, 017 (2001)
  [arXiv:hep-th/0012148].\\
  R.~Rattazzi and A.~Zaffaroni,
  JHEP {\bf 0104}, 021 (2001)
  [arXiv:hep-th/0012248].\\
  M.~Perez-Victoria,
  JHEP {\bf 0105}, 064 (2001)
  [arXiv:hep-th/0105048].


\bibitem{Agashe:2002jx}
  K.~Agashe and A.~Delgado,
  Phys.\ Rev.\  D {\bf 67}, 046003 (2003)
  [arXiv:hep-th/0209212].

\bibitem{Seiberg:1994pq}
  N.~Seiberg,
  Nucl.\ Phys.\  B {\bf 435}, 129 (1995)
  [arXiv:hep-th/9411149].

\bibitem{Harada:2003jx}
  M.~Harada and K.~Yamawaki,
  Phys.\ Rept.\  {\bf 381}, 1 (2003)
  [arXiv:hep-ph/0302103].

\bibitem{Cacciapaglia:2004jz}
  G.~Cacciapaglia, C.~Csaki, C.~Grojean and J.~Terning,
  Phys.\ Rev.\  D {\bf 70}, 075014 (2004)
  [arXiv:hep-ph/0401160].

\bibitem{:2005ema}
[LEP Collaborations],
  Phys.\ Rept.\  {\bf 427}, 257 (2006)
  [arXiv:hep-ex/0509008].

\bibitem{Peskin:1991sw}
  M.~E.~Peskin and T.~Takeuchi,
  Phys.\ Rev.\  D {\bf 46}, 381 (1992).

\bibitem{Sikivie:1980hm}
  P.~Sikivie, L.~Susskind, M.~B.~Voloshin and V.~I.~Zakharov,
  Nucl.\ Phys.\  B {\bf 173}, 189 (1980).

\bibitem{Csaki:2002gy}
  C.~Csaki, J.~Erlich and J.~Terning,
  Phys.\ Rev.\  D {\bf 66}, 064021 (2002)
  [arXiv:hep-ph/0203034].

\bibitem{Agashe:2003zs}
  K.~Agashe, A.~Delgado, M.~J.~May and R.~Sundrum,
  JHEP {\bf 0308}, 050 (2003)
  [arXiv:hep-ph/0308036].

\bibitem{Cacciapaglia:2004rb}
  G.~Cacciapaglia, C.~Csaki, C.~Grojean and J.~Terning,
  Phys.\ Rev.\  D {\bf 71}, 035015 (2005)
  [arXiv:hep-ph/0409126].

\bibitem{Agashe:2007ki}
  K.~Agashe {\it et al.},
  Phys.\ Rev.\  D {\bf 76}, 115015 (2007)
  [arXiv:0709.0007 [hep-ph]].

\bibitem{Lee:1977eg}
  B.~W.~Lee, C.~Quigg and H.~B.~Thacker,
  Phys.\ Rev.\  D {\bf 16}, 1519 (1977).

\bibitem{Gunion:1989we}
  J.~F.~Gunion, H.~E.~Haber, G.~L.~Kane and S.~Dawson,
  {\it The Higgs Hunter's Guide,} Westview Press (2000), 348pp.

\bibitem{Brodsky:2007hb}
  S.~J.~Brodsky and G.~F.~de Teramond,
  Phys.\ Rev.\  D {\bf 77}, 056007 (2008)
  [arXiv:0707.3859 [hep-ph]].

\bibitem{Hong:2004sa}
  S.~Hong, S.~Yoon and M.~J.~Strassler,
  JHEP {\bf 0604}, 003 (2006)
  [arXiv:hep-th/0409118].

\bibitem{Agashe:2005dk}
  K.~Agashe and R.~Contino,
  Nucl.\ Phys.\  B {\bf 742}, 59 (2006)
  [arXiv:hep-ph/0510164].

\bibitem{Konyushikhin:2009kq}
  M.~Konyushikhin,
  arXiv:0906.1904 [hep-ph].

\bibitem{Polchinski:2001tt}
  J.~Polchinski and M.~J.~Strassler,
  Phys.\ Rev.\ Lett.\  {\bf 88}, 031601 (2002)
  [arXiv:hep-th/0109174].

\bibitem{Alday:2007hr}
  L.~F.~Alday and J.~M.~Maldacena,
  JHEP {\bf 0706}, 064 (2007)
  [arXiv:0705.0303 [hep-th]].

\bibitem{McGreevy:2007kt}
  J.~McGreevy and A.~Sever,
  JHEP {\bf 0802}, 015 (2008)
  [arXiv:0710.0393 [hep-th]].

\bibitem{Polchinski:2002jw}
  J.~Polchinski and M.~J.~Strassler,
  JHEP {\bf 0305}, 012 (2003)
  [arXiv:hep-th/0209211].

\bibitem{Early Composite Pheno}
  T.~G.~Rizzo,
  Phys.\ Rev.\  D {\bf 32}, 43 (1985).\\
  M.~Suzuki,
  Phys.\ Lett.\  B {\bf 153}, 289 (1985).\\
  J.~A.~Robinson and T.~G.~Rizzo,
  Phys.\ Rev.\  D {\bf 33}, 2608 (1986).\\
  G.~Couture and J.~N.~Ng,
  Z.\ Phys.\  C {\bf 32}, 579 (1986)
  [Erratum-ibid.\  C {\bf 43}, 522 (1989)].

\bibitem{Hagiwara:1986vm}
  K.~Hagiwara, R.~D.~Peccei, D.~Zeppenfeld and K.~Hikasa,
  Nucl.\ Phys.\  B {\bf 282}, 253 (1987).

\bibitem{Hagiwara:1989mx}
  K.~Hagiwara, J.~Woodside and D.~Zeppenfeld,
  Phys.\ Rev.\  D {\bf 41}, 2113 (1990).

\bibitem{Alcaraz:2006mx}
  J.~Alcaraz {\it et al.}  [LEP Collaborations],
  arXiv:hep-ex/0612034.

\bibitem{Iashvili:2008dj}
  I.~Iashvili  [CDF and D0 Collaborations],
  arXiv:0809.2955 [hep-ex].

\bibitem{ATLAS TDR}
ATLAS Collaboration,
vol.\ II, CERN-LHCC-99-15, ATLAS-TDR-15 (May 1999).

\bibitem{VVjj Refs}
  R.~N.~Cahn, S.~D.~Ellis, R.~Kleiss and W.~J.~Stirling,
  Phys.\ Rev.\  D {\bf 35}, 1626 (1987).\\
  R.~Kleiss and W.~J.~Stirling,
  Phys.\ Lett.\  B {\bf 200}, 193 (1988).\\
  V.~D.~Barger, T.~Han and R.~J.~N.~Phillips,
  Phys.\ Rev.\  D {\bf 37}, 2005 (1988).\\
  V.~D.~Barger, K.~m.~Cheung, T.~Han and D.~Zeppenfeld,
  Phys.\ Rev.\  D {\bf 44}, 2701 (1991)
  [Erratum-ibid.\  D {\bf 48}, 5444 (1993)].

\bibitem{AtlasBook}
  G.~Aad {\it et al.}  [The ATLAS Collaboration],
  arXiv:0901.0512 [hep-ex].

\bibitem{Brig:2007}
  V.~Brigljevic {\it et al.},
  J.\ Phys.\ G {\bf 34}, N269 (2007).

\bibitem{Rizzo:2006nw}
  T.~G.~Rizzo,
  arXiv:hep-ph/0610104.

\bibitem{Kumar:2006gm}
  J.~Kumar and J.~D.~Wells,
  Phys.\ Rev.\  D {\bf 74}, 115017 (2006)
  [arXiv:hep-ph/0606183].

\bibitem{Pomarol:1999ad}
  A.~Pomarol,
  Phys.\ Lett.\  B {\bf 486}, 153 (2000)
  [arXiv:hep-ph/9911294].

\bibitem{Sakurai:1960ju}
  J.~J.~Sakurai,
  Annals Phys.\  {\bf 11}, 1 (1960).

\end{thebibliography}
\end{document}